\documentclass{emulateapj}
\usepackage{natbib}
\usepackage[bookmarks=false, colorlinks=true, citecolor=blue,
  backref=true]{hyperref}              
\hypersetup{
  colorlinks,
  citecolor=black,
  linkcolor=red,
  urlcolor=blue,
}

\newcommand{\um}{\mbox{$\,\mu{\rm m}$}}
\newcommand{\mum}{\mbox{$\,\mu{\rm m}$}}

\newcommand{\kms} {\mbox{\,km~s$^{-1}$}}

\newcommand{\Herschel}{{\it Herschel}}

\newcommand{\FeII}{\mbox{[Fe\,{\sc ii}]}}
\newcommand{\SiVI}{\mbox{[Si\,{\sc vi}]}}
\def\co#1#2{{\hbox {${\mathrm{CO}}\,(#1\text{--}#2)$}}}
\def\cs#1#2{{\hbox {${\mathrm{CS}}\,(#1\text{--}#2)$}}}

\shorttitle{ALMA Observation of NGC\,5135}
\shortauthors{Cao et al.}

\begin{document}

\slugcomment{{\bf Accepted for publication in APJ}; \today}

\title{ALMA Observation of NGC\,5135: The Circumnuclear \co65\ and Dust Continuum 
       Emission at 45 Parsec Resolution\footnotemark[$\star$]}
\author{
Tianwen Cao \altaffilmark{1,2,3,4},
Nanyao Lu \altaffilmark{1},
C. Kevin Xu \altaffilmark{1},
Yinghe Zhao \altaffilmark{5,6,7},
Venu Madhav Kalari \altaffilmark{8},
Yu Gao \altaffilmark{9},
Vassilis Charmandaris \altaffilmark{10,11},
Tanio Diaz Santos \altaffilmark{12},
Paul van der Werf \altaffilmark{13},
Chen Cao \altaffilmark{14,15},
Hong Wu \altaffilmark{2,3},
Hanae Inami \altaffilmark{16},
Aaron Evans \altaffilmark{17,18}
}

\altaffiltext{1}{Chinese Academy of Sciences South America Center
  for Astronomy, National Astronomical Observatories,
  Chinese Academy of Sciences, Beijing 100101, China;
  twcao@bao.ac.cn}
\altaffiltext{2}{Key Laboratory of Optical Astronomy,National Astronomical Observatories, Chinese Academy of Sciences, Beijing 100101, China}
\altaffiltext{3}{School of Astronomy and Space Science, University of Chinese Academy of Sciences, Beijing 100049, China}
\altaffiltext{4}{Instituto de Astrofisica, Pontificia Universidad  Cat\'olica de Chile,$\!$Av.$\!$Vicu\~na Mackenna$\!$4860,\,7820436 Macul,$\!$Santiago, Chile}
\altaffiltext{5}{Yunnan Observatories, Chinese Academy of Sciences, Kunming 650011, China}
\altaffiltext{6}{Key Laboratory for the Structure and Evolution of Celestial Objects, Chinese Academy of Sciences,Kunming 650011, China}
\altaffiltext{7}{Center for Astronomical Mega-Science, CAS, 20A Datun Road, Chaoyang District, Beijing 100012, China}
\altaffiltext{8}{Departamento de Astronomía, Universidad de Chile, Casilla 36-D, Correo Central,Santiago,Chile}
\altaffiltext{9}{Purple Mountain Observatory/Key Lab of Radio Astronomy, 8 YuanHua Road, 210034 Nanjing, China}
\altaffiltext{10}{Department of Physics, University of Crete, GR-71003 Her-
  aklion, Greece}
\altaffiltext{11}{IAASARS, National Observatory of Athens, GR-15236, Pen-teli, Greece}
\altaffiltext{12}{Núcleo de Astronomía de $\!$la$\!$Facultad $\!$de$\!$Ingeniería, Universidad Diego Portales,Av. Ejercito Libertador 441,Santiago,Chile}
\altaffiltext{13}{Leiden Observatory, Leiden University, P.O. Box 9513, 2300 RA Leiden, The Netherlands}
\altaffiltext{14}{School of Space Science and Physics, Shandong University,
  Weihai, Weihai, Shandong 264209, China}
\altaffiltext{15}{Shandong Provincial Key Laboratory of Optical Astronomy \& Solar-Terrestrial$\!$Environment,$\!$ Weihai,$\!$ Shandong $\!$264209,$\!$China}
\altaffiltext{16}{Univ. Lyon, Univ. Lyon1, ENS de Lyon, CNRS, Centre de Recherche Astrophysique de Lyon (CRAL) UMR 5574, 69230 Saint-Genis-Laval, France}
\altaffiltext{17}{Department of Astronomy, University of Virginia, Charlottesville, VA 22903-2325, USA}
\altaffiltext{18}{National Radio Astronomy Observatory, 520 Edgemont Road, Charlottesville, VA 22903-2475, USA}

\footnotetext[$\star$]{
The National Radio Astronomy Observatory is a facility of the National Science Foundation operated under cooperative  
agreement by Associated Universities, Inc.}


\begin{abstract}
We present high-resolution (0.17\arcsec $\times$ 0.14\arcsec)
Atacama Large Millimeter/submillimeter Array (ALMA) observations
of the \co65\ line, and 
435\um\ dust continuum emission within a $\sim$9\arcsec $\times$ 9\arcsec\ 
area centered on the nucleus of the galaxy NGC\,5135.
NGC\,5135 is a well-studied luminous infrared 
galaxy that also harbors a Compton-thick active galactic nucleus (AGN).  
At the achieved resolution of 48 $\times$ 40\,pc, the \co65\
and dust emissions are resolved into gas ``clumps'' along the symmetrical dust
lanes associated with the inner stellar bar. 
The clumps have radii between $\sim$45-180\,pc and  
\co65\ line widths of $\sim$60-88\,\kms. The \co65\ to dust 
continuum flux ratios vary among the clumps and show 
an increasing trend with the \FeII/Br-$\gamma$ ratios, which 
we interpret as evidence for supernova-driven shocked gas providing a significant contribution to the \co65\ emission. 
The central AGN is undetected in continuum, nor in CO (6-5) if its line 
velocity width is no less than $\sim$\,40\,\kms.
We estimate that the AGN contributes at most 1\% of the 
integrated \co65\ flux of 512 $\pm$ 24$\,$Jy\kms\
within the ALMA field of view, 
which in turn accounts for $\sim$32\% of the \co65\ flux of the whole galaxy.
\end{abstract}

\keywords{galaxies: active -- galaxies: starbusrt -- galaxies: general -- galaxies: nuclei -- galaxies: ISM}

\section{Introduction \label{intro}}

Luminous infrared galaxies (LIRGs; $L_{\rm IR}^{[8-1000\,\mu\rm m]} \gtrsim 
10^{11}\,L_{\odot}$), whose space density exceeds that 
of optically selected starburst and Active Galactic Nucleus (AGN)
host galaxies at comparable bolometric luminosities 
\citep{1987ApJ...320..238S}, consist of isolated galaxies,
galaxy pairs, interacting galaxy systems and advanced mergers 
\citep{1996ARA&A..34..749S,2006ApJ...649..722W}. LIRGs in the later
stages of evolution tend to
contain a rich amount of molecular gas in the galaxy nuclear 
region \citep{1986ApJ...305L..45S}, and have a higher fraction 
of AGNs compared with less luminous galaxies \citep{1996ARA&A..34..749S}.
Detailed investigations of the physical properties, AGN-starburst 
connection, and gas inflow/outflow in representative LIRGs in 
the local Universe are critical to our understanding of galaxy 
evolution because LIRGs are the dominant contributors to the cosmic 
star formation (SF) at $z \gtrsim 1$ (\citealt{2005ApJ...632..169L, 
2007ApJ...660...97C, 2009A&A...496...57M, 2011A&A...528A..35M, 
2013MNRAS.432...23G})

The CO emission lines from low-$J$ transitions, such as CO\,(1-0) 
 at 2.6\,mm and CO\,(2-1) at 1.3\,mm, have 
been widely used to trace the molecular gas content in LIRGs 
\citep{1988ApJ...334..613S, 1991ApJ...370..158S, 
1997ApJ...478..144S, 1999AJ....117.2632B, 
1999ApJ...512L..99G, 2002ApJ...580..749E}. 
However, based on the data taken with the SPIRE Fourier Transform
Spectrometer (FTS; \citealt{2010A&A...518L...3G}) onboard 
the {\it Herschel Space Observatory} ({\it Herschel}; 
\citealt{2010A&A...518L...1P})
on a flux-limited sample of 123 LIRGs from the Great Observatories 
All-Sky LIRGs Survey (GOALS; \citealt{2009PASP..121..559A}),
\citet{2014ApJ...787L..23L, 2017ApJS..230....1L} showed that 
the mid-$J$ CO emission (i.e., 4 $<J<$ 10) from warm and dense
molecular gas correlates linearly with the star formation rate (SFR) 
on galactic scale for LIRGs over a wide range of $L_{\rm IR}$ and 
far-infrared (FIR) color. Therefore, the heating mechanism for 
the warm dense gas that gives rise to the mid-$J$ CO line emission
should ultimately derive the energy from the same SF process that 
powers the dust emission. There is not yet a firm consensus 
on this heating mechanism. Apparently different heating mechanisms 
are favored from analyses of the CO emission line spectra of 
individual galaxies, including far-UV photon heating (e.g., 
\citealt{2013MNRAS.434.2051R}), heating by cosmic rays enhanced 
by supernovae (SNe; e.g., \citealt{2003ApJ...586..891B}), and 
heating by shocks that may or may not be powered by SNe
(e.g., \citealt{2012ApJ...753...70K,
2011ApJ...743...94R, 2011ApJ...742...88N, 2013ApJ...762L..16M,
2013ApJ...779L..19P, 2014A&A...568A..90R}).
The X-ray photons from an AGN can heat the surrounding dense 
gas very effectively (e.g., \citealt{2008ApJ...678L...5S}).
However, \citet{2017ApJS..230....1L} argued that the CO line 
emission associated with any AGN gas heating may peak at
$J > 10$. As a result, the mid-$J$ CO line emission is  
always dominated by SF.

With the Atacama Large Millimeter/submillimeter Array (ALMA; 
\citealt{2009IEEEP..97.1463W}), it is now possible to obtain 
high-resolution mid-$J$ CO line and dust continuum images of 
the nuclei of nearby LIRGs to investigate whether the {\it Herschel} 
results above still hold true at physical scales down to 
the typical size of giant molecular clouds (GMCs; i.e., $\sim$40 pc, 
 \citealt{2009ApJS..184....1K}), and whether one can rule out some 
of the gas heating mechanisms proposed. To this end, we have 
carried out a number of ALMA Band-9 observations over time, to
image simultaneously the CO\,(6-5) line emission (the rest frequency 
$\nu_{\rm rest}$ = 691.473\,GHz) and its underlying dust continuum at 
435\um\ in the nuclear regions of a set of carefully selected,
representative LIRGs from our {\it Herschel} FTS sample.
The targets observed include NGC\,34 \citep{2014ApJ...787...48X}
and NGC\,1614 \citep{2015ApJ...799...11X}, two advanced mergers
with a warm FIR color;  NGC\,7130 \citep{2016ApJ...820..118Z}
and NGC\,5135 (in this paper), two well-known Seyfert galaxies
with a prominent stellar bar; IC\,5179 \citep{2017ApJ...845...58Z},
an isolated, unbarred galaxy with a compact nuclear starburst; 
and CGCG\,049-057 with a high-surface density nuclear SF disk 
(Cao et al., in preparation). The linear resolutions ($R_{\rm linear}$)
achieved range from 100\,pc in the early observation of 
NGC\,34 to 34\,pc in the case of IC\,5179. Three additional 
LIRGs in our FTS sample also have ALMA \co65\ images in 
the literature: Arp\,220 ($R_{\rm linear} \sim 165$\,pc;  
\citealt{2014ApJ...789L..36W, 2015ApJ...806...17R}),   
IRAS 13120-5453 ($R_{\rm linear} \sim 165$ pc; \citealt{2017ApJ...840L..11S}),  
and NGC\,1068 ($R_{\rm linear} \sim 4$\,pc; \citealt{2014A&A...567A.125G, 
2016ApJ...823L..12G}).  ALMA CO\,(6-5) images also exist for 
two nearby, but non-LIRG galaxies: NGC\,1377 \citep{2017A&A...608A..22A}
and Centaurus\,A \citep{2017ApJ...843..136E}.

At a distance of 59\,Mpc (1\arcsec\ corresponds to 281\,pc) and 
with a fairly face-on disk,  NGC\,5135 is a well studied LIRG 
with $L_{\rm IR} = 10^{11.33}\,L_{\odot}$ and a moderately warm 
FIR color of 0.54 (in terms of the 60-to-100\,\um\ flux density 
ratio; \citealt{2009PASP..121..559A}). The galaxy not only displays
a powerful circumnuclear starburst over a region of $\sim$1\,kpc 
in diameter \citep{1998ApJ...505..174G, 2009ApJ...698.1852B} 
but also harbors a highly obscured Seyfert\,2 nucleus 
\citep{2004ApJ...602..135L, 1983ApJ...266..485P, 1997ApJS..113...23T}. 
It is therefore an ideal target for high-resolution ALMA observations
to separate the circumnuclear SF from the AGN. The 6\,cm radio 
continuum emission peaks in an area $\sim$3\arcsec\ south of
the nucleus based on a Very Large Array (VLA) observation by 
\citet{1989ApJ...343..659U}, presumably tracing the supernova 
remnants (SNRs) from a previous starburst. The high-resolution {\it 
Hubble Space Telescope} ({\it HST}) UV/optical imaging observations 
unveiled a large number of young star clusters, between the nucleus
and the radio continuum peak \citep{1998ApJ...505..174G}, which
presumably have partially cleared gas. Furthermore, the high-resolution 
near- and mid-infrared images \citep{2006ApJ...650..835A, 
2008ApJ...685..211D} show patches of strong on-going SF along, 
but at the downstream side of the dust lanes that are likely 
associated with the stellar bar (e.g. \citep{1997ApJ...482L.135M}).  
An intermediate resolution ($R 
\sim$ 3000$-$4000), near-infrared integral-field spectroscopy 
\citep{2009ApJ...698.1852B} confirmed the presence of a high-excitation 
ionization cone centered on the AGN, based on the \SiVI\ 1.96\um\ 
line emission, as well as an extended distribution of shocked gas 
likely powered by SNe, based on the \FeII\ 1.46\um\ line.  
\citet{2011ApJ...727...19F} and \citet{2012MNRAS.419.2089S} obtained
broad-band (10$-$50\,kev) X-ray spectra of NGC\,5135, demonstrating 
that the AGN in NGC\,5135 is obscured by Compton-thick material. 
Our ALMA imaging of NGC\,5135 presented here provides for the first time 
the distribution and kinematics of the warm and dense molecular gas
as well as the morphology of the 435\um\ dust emission, at a linear 
resolution of less than 50\,pc in the circumnuclear region of NGC\,5135.

In the remainder of the paper, we describe our ALMA observation
and data reduction in \S2 and present our results in \S3. In \S4, 
we discuss the physical implications derived from our data on the circumnuclear 
SF and the role of the AGN, compare our ALMA images with existing 
images at other wavelengths, and comment on the most likely heating
mechanism for the observed \co65\ emission.
Thereby allowing us to distinguish between different heating mechanisms,
and the role played by SF in giving rise to the mid-J CO emission.
Finally, we summarize our results in \S5. Throughout this paper, 
we adopt a distance of 59\,Mpc for NGC\,5135 (Armus et al. 2009).

\section{OBSERVATION AND DATA REDUCTION \label{obsxdar}}

The ALMA band-9 observation of NGC\,5135 was carried out in the time 
division mode (with a velocity resolution of $\sim$6.8\,\kms). 
The four basebands (i.e., Spectral Windows; SPWs 0-3) were centered 
on sky frequencies of 681.975, 683.736, 678.243 and 680.183\,GHz, 
respectively, with a bandwidth of 1.875\,GHz. The observation was 
performed with the configuration mode C34-5, using 39 12-meter 
antennae with baselines ranging from 21.3 to 885.6 meters.
The total on-target integration time is 21.03 minutes.  
During the observation, the phase calibration and amplitude were monitored
 using J1316-3328. 
Additional observing details can been found in Table\,1.

The data were reduced using the Common Astronomical Software
Application (CASA) version 4.5 \citep{2007ASPC..376..127M}.   
Our primary beam is $\sim$8.8$''$ and 
the Maximum Recoverable Scale (MRS)\footnotemark[1] is 3.5\arcsec.
The \co65\ line data cube was generated using the data in the SPW 
covering the sky frequency range of 680.975 to 682.975\,GHz. 
The continuum was estimated 
by combining the data from the other three SPWs. The calibrated images 
were cleaned using the Briggs weighting (with the parameter 
``robust'' set to 0.5). The resulting synthesized beams are 
nearly identical between the continuum and the line emission 
and have a full width at half maximum (FWHM) size of $\sim$ 
0.17\arcsec\,$\times$\,0.14\arcsec (equivalent to 48 $\times$
40\,pc at the distance to NGC\,5135), with the major-axis position   
angle (north to east) at 111\arcdeg.  All our analyses in 
this paper use the data after the primary beam correction, whereas 
all of the figures (except Fig.\,6) were produced using data prior to the primary 
beam correction.
\footnotetext[1]{https://science.nrao.edu/science/videos/largest-angular-scale-and-maximum-recoverable-scale}

The final spectral cube has a channel width equivalent to 13.5\kms\
in velocity. The channel noise ($\sigma_{\rm ch}$) is on the order 
of 18$\,$mJy\,beam$^{-1}$.  
The total CO\,(6-5) flux image, as an integration over 
the barycentric velocities from 3,971 to 4,157\kms,  has 
an r.m.s.~noise of $\sim$1.2 Jy\,beam$^{-1}$\kms. 
The continuum image has a noise of 2.2\,mJy\,beam$^{-1}$.  
All these noise measurements were done on the data without 
the primary beam correction.
The ALMA absolute flux calibration is estimated to be good 
to $\sim$10\%. The astrometric accuracy is better than 0.01\arcsec.

\section{Results \label{re}}

\subsection{{\rm CO\,(6-5)} Line Emission \label{COLE}}

\begin{figure*}[h!tb]
\begin{center} 
\includegraphics[width=7.0in]{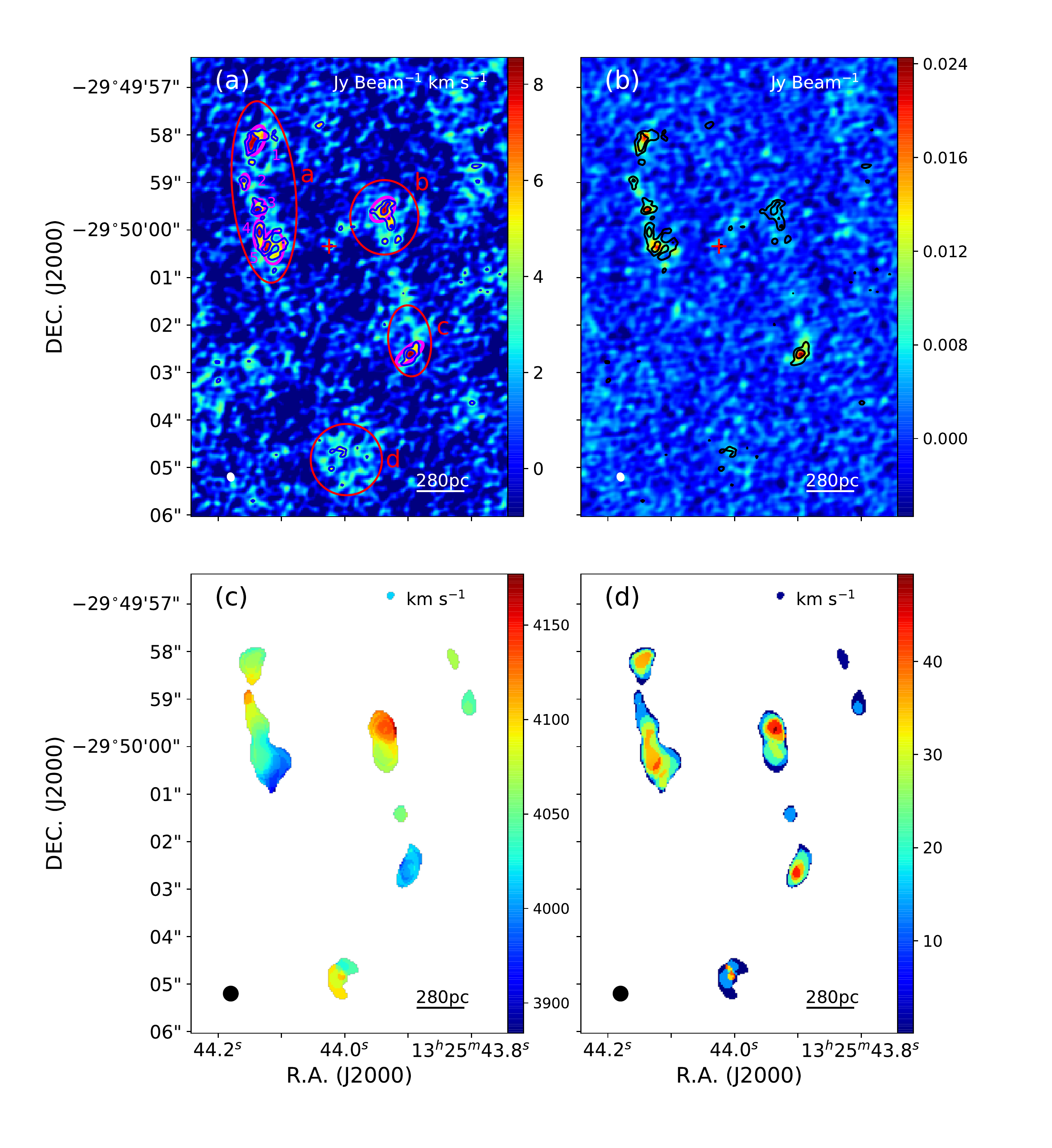}
\caption{
Panels (a) and (b): Contours of the frequency-integrated CO\,(6-5) 
intensity overlaid on the images of the same integrated CO\,(6-5) 
intensity in (a) and the 435\um\ dust continuum emission in (b).
Panels (c) and (d) are respectively the line velocity field (moment 1)
and the velocity dispersion (moment 2) maps of the CO\,(6-5) line
emission obtained from the uv-taper image. The images in (c) and (d) 
are generated by using only those spaxels 
above 4-$\sigma_{ch}$, where $\sigma_{ch}$ is the r.m.s noise per frequency 
channel ($\sigma_{ch}$ = 40\,mJy\,beam$^{-1}$ for uv-taper image).
The contours in panels (a) and (b) are shown at [3,5]$\times 
\sigma$ (where the noise $\sigma$ = 1.2\,Jy\,beam$^{-1}$\,kms$^{-1}$). 
The unit of the color bar in each panel is given near the upper-right 
color. The filed ellipse in white near the lower-left corner in (a) or 
(b) is the ALMA beam. The large ellipses in red in (a) mark the regions 
for spectrum and flux extractions given in Table\,2.  The red plus sign 
marks the AGN position adopted.
The figures are before the primary beam correction integrated over 
the barycentric velocities from 3,971 to 4,157\kms.
\label{figure1}}
\end{center}
\end{figure*}

The four panels in Fig.\,\ref{figure1} show respectively the images 
of the total CO\,(6-5) emission integrated between the observed 
velocities of 3971 and 4157\,\kms, the 435\,\um\ continuum, 
the velocity field (i.e., moment 1), and velocity dispersion map
 (moment 2). Each image roughly covers the ALMA primary 
beam. The contours overlaid in Fig.\,1a and 1b refer 
to the same total CO\,(6-5) emission and start at S/N $= 3$.

The CO\,(6-5) emission detected at S/N $> 3$ appears clumpy and
is confined to a few discrete regions along two spiral arm-like
features corresponding spatially to the dust lanes seen
in the UV and optical (e.g., \citealt{2007AJ....134..648M}). 
We mark four separate regions, namely ({\it a}, {\it b}, {\it c}, 
{\it d}) which are compact and concentrated in the intergrated
  CO\,(6-5) map.
The region {\it d} appears more diffuse 
compared with the other three. The total CO\,(6-5) flux from
the combined four regions in Fig.\,1a is 512 $\pm$ 24\,Jy\kms.  
  With a much larger beam of $\sim$31\arcsec, the {\Herschel}/FTS 
observation gives a \co65\ flux of 1,617\,Jy\kms 
\citep{2017ApJS..230....1L}. Therefore the clump regions in 
Fig.\,1a together account for $\sim$32\% of the total CO\,(6-5) 
flux of the galaxy.  The $''$missing$''$ line flux could be due to 
a combination of the line emission outside the ALMA field of 
view, or possible faint emission at peak surface brightness 
below our 3-$\sigma$ (i.e., 3.6 Jy\,beam$^{-1}$\,\kms) detection
limit, or resolved out on larger scales.
We analyze individual clumps in more detail in \S4.2.

\begin{figure*}[!htbp]
\begin{center}
\includegraphics[width=8.4in]{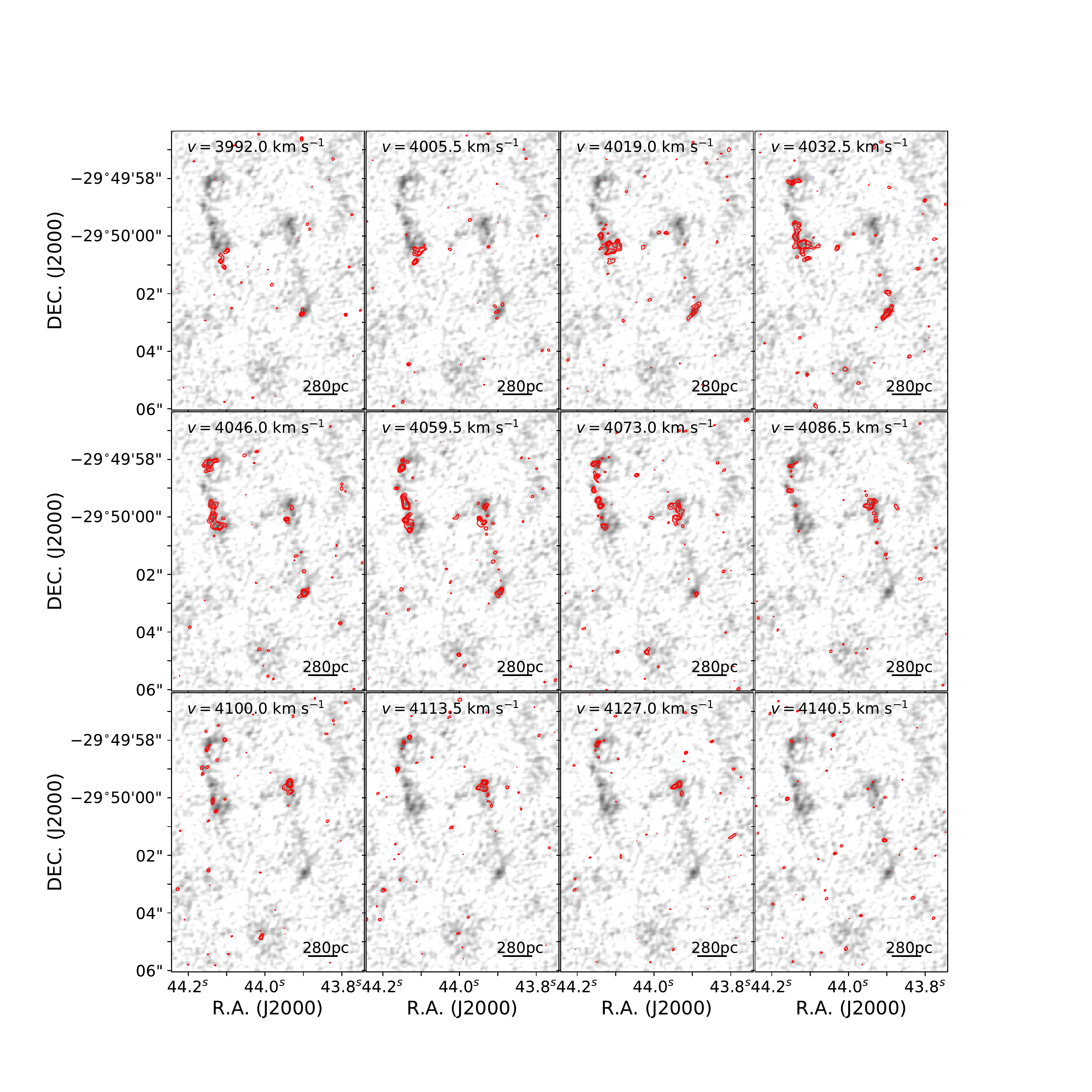}
\caption{ 
Channel maps of the CO\,(6-5) line emission (in contours), 
each overlaid on the image of the total, frequency-integrated 
CO\,(6-5) emission (e.g., from Fig.\,1a). The channel interval 
is 13.5\kms, with the channel central (barycentric)
velocity shown in each channel map. 
\label{channelmap}}
\end{center}
\end{figure*}

We set the threshold at 4-$\sigma_{\rm ch}$,
where $\sigma_{\rm ch}$ is the r.m.s noise per frequency channel,
 to obtain the moment 1 and 2 maps. To reveal the kinematics better,
 the moment 1 and 2 maps, shown respectively in Fig.\,1c and 1d, were based on the uv-taper image (the details about the uv-taper image are presented in the last paragraph of this subsection).
The velocity scale in Fig.\,\ref{figure1}c was calculated using 
the formula $\nu_{\rm obs} = \nu_{\rm rest}\,(1 - V/c)$, where 
$\nu_{\rm obs}$ is the observed CO\,(6-5) line frequency, $c$ 
the speed of light, and $V$ the velocity to calculate.
The line velocity ranges from 3992 to 4140\,\kms.  
Fig.\,\ref{figure1}d shows that the line-of-sight velocity dispersion
ranges from 10 to 40\,\kms, using only those pixels with S/N
$>$ 4-$\sigma_{\rm ch}$.
The overall kinematic pattern can also be seen 
in the channel maps displayed in Fig.\,\ref{channelmap}, 
where the contours from an individual channel of width
13.5\,\kms are overlaid on the grayscale image of the total 
CO\,(6-5) flux map shown in Fig.\,1a. 
While the regions {\it a, c, d} are mainly confined within a velocity 
range of 3992 to 4073\,\kms, the region {\it b} has a range 
between 4046 and 4127\,\kms.
This suggests that the observed velocity pattern is not
dominated by a simple rotation within the galaxy disk.
In the channel maps, some CO\,(6-5) clumps break down into 
smaller clumps (or clouds) in some velocity channels, e.g., 
the region {\it a} in the channel centered at $V = 4073$\kms.  
These clouds have sizes $< 50$\,pc.

In Fig.\,3a and 3b, we reproduced the same images as in Fig.\,1a
and 1b, respectively, but with a larger effective beam 
of 0.4$''$ $\times$ 0.4$''$ (equivalent to 112 $\times$ 112 pc)
by applying an uv-taper (with 
the parameter ``outertaper'' = 0.4$''$) to our uv data before
imaging. The peak signal of the region {\it d} is higher than 
4-$\sigma$ ($\sigma$ = 8\,Jy\,beam$^{-1}$\,\kms). The total flux 
of the four regions combined, as defined in Fig.\,\ref{figure1}, 
is 841 $\pm$ 45\,Jy\kms. This flux equals 1.6 times the flux 
from the original ALMA image, and is $\sim$52$\%$ of the total 
flux measured by {\it Herschel}, confirming 
that there exists some more diffuse or lower surface brightness 
CO\,(6-5) emission within the region of Fig.\,1a.

\subsection{Dust Continuum Emission \label{CONT}}

As shown in Fig.\,3b, the continuum at 435\um\ generally coincides
with the CO\,(6-5) line emission in regions {\it a} and {\it c}
at scales of 0.4\arcsec\ (equivalent to 112\,pc), which 
corresponds to an angular size of 114\,pc at the distance of 
NGC\,5135. This is consistent with the findings in
the other LIRGs we imaged in CO\,(6-5), i.e., NGC\,34, NGC\,1614 
and NGC\,7130 and IC\,5179 \citep{2014ApJ...787...48X, 2015ApJ...799...11X,  
2016ApJ...820..118Z, 2017ApJ...845...58Z}, i.e., at scales $\gtrsim$ 
100\,pc, there is a good spatial correspondence between 
the CO\,(6-5) line and its underlying continuum emissions.

However, at scales significantly smaller than 100\,pc, there 
are apparent offsets between the local peaks of 
the line and continuum emissions in Fig.\,1b. This small-scale
offset between the line and continuum emissions is also seen 
in the LIRGs of moderately high nuclear gas surface densities, 
e.g., IC\,5179 (at linear resolution $R_{\rm linear} \approx
34$\,pc; \citealt{2017ApJ...845...58Z}) and NGC\,7130 
($R_{\rm linear} \approx 70$\,pc $\times$ 40\,pc; \citealt{2016ApJ...820..118Z}).
Furthermore, in both Fig.\,1b and 3b, the dust continuum is 
unusually weak relative to the line emission in the regions 
{\it b} and {\it d}. As argued in \citet{2016ApJ...820..118Z}, 
these differences between the line and dust continuum emissions 
at small scales can only be understood if the gas and dust are 
heated by different mechanisms. We discuss this in more detail 
in \S4.3.

By combining the four regions in Fig.\,1b, we derived a total 
flux of 181 $\pm$ 25\,mJy for the 435\um\ continuum emission. 
This flux would be 1.6 times higher if we had derived it from 
the same regions in Fig.\,3b.

\section{ANALYSIS AND DISCUSSION \label{A&D}}

\subsection{The Central AGN \label{AR}}

The AGN position can be constrained by the peak of the \SiVI\ 
line emission at 1.96\,\um\ in a ground-based observation  
\citep{2009ApJ...698.1852B} and by the peak of the hard X-ray 
(4-8 keV) emission detected with Chandra \citep{2004ApJ...602..135L}.
The estimated astrometric uncertainty associated 
with either of these images is on the order of 0.5\arcsec. 
The VLA 6\,cm radio continuum image of \citet{1989ApJ...343..659U} 
has a modest resolution of 0.91\arcsec\ $\times$ 0.60\arcsec\ and
an astrometric accuracy of 0.3\arcsec. We overlaid our CO\,(6-5) 
contours from Fig.\,1a on this radio continuum image in Fig.\,4a.
As already stated in Sec.\,1, the main peak of the 6\,cm emission 
is $\sim$3\arcsec\ south of the galaxy nucleus, spatially coincident
with the peak of the broad (FWHM = $\sim$513\kms)
[Fe{\scriptsize{II}}] emission \citep{2009ApJ...698.1852B}.
However, there is a minor radio emission peak (10$\sigma$) 
near the anticipated AGN position.  
We take the position of this radio peak 
(R.A.~= 13$^h$25$^m$44$^s$.02, Dec.~= $-$29$^{\circ}$50$'$00$''$.4; J2000) 
as the AGN location (i.e., marked by the white cross), with a positional
uncertainty of 0.3-0.5\arcsec. In Fig.\,4b, we overlaid the same 
CO\,(6-5) contours on the hard X-ray emission.

As shown in Fig.\,1a, the \co65\ emission is
undetected at 4-$\sigma$ level at the AGN 
position (i.e., integrated over the velocity 
range of 186\,\kms). This would hold true 
even if we had lowered the detection threshold 
to 3-$\sigma$.
We assume the AGN-related \co65\ emission is confined to an area smaller than 
our ALMA beam size ($\sim$48\,$\times$\,40\,pc), then the 3-$\sigma$ 
flux upper limit is equal to 3\,$\times$\,(1.2\,Jy\,\kms) = 3.6\,Jy\,\kms 
($\sigma$ = 1.2\,Jy\,beam$^{-1}\,$\kms).

However, an apparent narrow emission feature at the AGN 
location is seen at 3-4 $\sigma$ significance over only two 
velocity channels (i.e., V = 4019.0 and 4032.5\,\kms).  
The image summed over these two velocity channels is presented
 in Fig.\,5a, which shows a peak surface brightness of 33 mJy\,beam$^{-1}$
(at 5-$\sigma$ significance). The spectrum in Fig.\,5b is 
extracted from a circular aperture of radius = 0.4$''$ 
(= 2.5 times the FWHM of the 3-$\sigma$ surface brightness 
of the emission in Fig.\,5a). Its narrow velocity width 
of $\sim$40\,\kms makes it unlikely that this signal is physically 
associated with the AGN. Nevertheless, considering 
that the \co65\ emission associated with the gas torus 
of the AGN in NGC\,1068 is observed to have only a 
modest velocity width of $\sim$80\,\kms \citep{2016ApJ...823L..12G},
we defer to a future observation of higher angular resolution
to firmly conclude the reality of this narrow \co65\ emission.  
Flux-wise, the narrow \co65\ emission in Fig.\,5b has a flux 
of 1.7\,Jy\,\kms, which is smaller than the 3-$\sigma$ 
flux upper limit of 3.6\,Jy\,\kms derived above. Therefore, 
we conclude that the AGN in NGC\,5135 contributes at most 
1\% of the \co65\ flux observed within the ALMA field of view. 
This is consistent with the {\Herschel} finding that the mid-$J$
CO line emission in LIRGs is mainly associated with SF 
regardless of whether there is an AGN or not \citep{2017ApJS..230....1L}.  
The fractional contribution of the AGN to the
bolometric luminosity of NGC\,5135 is about 
(24$\pm$6)\% \citep{2017ApJ...846...32D}.
The AGN in NGC\,5135 is heavily obscured; the surrounding gas 
could be heated to a very high temperature by the X-rays associated
with the AGN, resulting in a CO spectral line distribution 
that peaks $J > 10$ \citep{2008ApJ...678L...5S}. Such a scenario
seems to be the case in the Seyfert galaxy NGC\,1068: While 
the high-resolution ALMA imaging shows that the vast majority 
of the nuclear CO\,(6-5) emission is associated with the compact
circumnuclear ring of SF at a radius of $\sim$100\,pc 
\citep{2016ApJ...823L..12G}, the total nuclear CO emission line
spectrum has a distinct component that peaks at $J \sim 16$ 
\citep{2012ApJ...758..108S}. This hot spectral component of 
the CO emission is presumably due to the AGN in NGC\,1068. 
The central AGN in NGC\,5135 is bright in terms of the 1-0\,S(1) 2.12\,\um\ 
H$_2$ ro-vibrational line \citep{2009ApJ...698.1852B} associated
with warm molecular gas. Although this line could be excited by 
different physical processes, including UV-fluorescence (photons); 
shock fronts (collisions) and X-ray illumination, \citet{2009ApJ...698.1852B} 
argued that the excited near-IR H$_2$ emission is mainly caused 
by X-ray illumination in the AGN region of NGC\,5135. Such an
X-ray dominant scenario is also favored based on  
non-detection of the \co65\ emission here.

The 435\um\ dust continuum is also undetected at the AGN position, 
with the 3-$\sigma$ flux upper limit equal to 
5.4\,mJy ($\sigma$ = 1.8\,mJy\,beam$^{-1}$; the same method 
used for deriving the \co65\ flux upper limit).
We compare this flux upper limit with the expected 435\um\ 
continuum flux from an average infrared spectral energy 
distribution (SED) appropriate for the AGNs of X-ray 
luminosities comparable to that of NGC\,5135: The intrinsic
2.0-10 keV X-ray luminosity of NGC\,5135 is $\sim$1.8 
$\times$ 10$^{43}$ erg\,s$^{-1}$ \citep{2012MNRAS.419.2089S}. 
We therefore used the infrared AGN SED for $L_{2.0-10\,\,\rm keV} 
> 10^{42.9}$ erg\,s$^{-1}$ in \citet{2011MNRAS.414.1082M} 
and anchored it at the 12\um\
luminosity of NGC\,5135 estimated from the X-ray and mid-IR 
correlation given in \citet{2015MNRAS.454..766A}. This
derived SED is shown in Fig.\,\ref{agntem}, along with two
continuum flux upper limits (at 3$\sigma$) at 435 and 1,300\um\ 
based on the ALMA observation. The latter continuum flux 
upper limit was estimated from an archival ALMA Band-6 
observation (Project 2013.1.00243.S; PI: L. Colina).  
 This plot suggests that the ALMA data points are consistent 
with what is expected from the typical infrared SED for AGNs like NGC\,5135.

\subsection{Properties of Molecular Gas Clumps \label{clu}}

Several clumpy features are resolved in the \co65\ image shown in Fig.\,1a. The resolved clump sizes of $\sim$100 pc are comparable with or larger than the beam size.
 For other (U)LIRGS, such clumpy features are traced more commonly by low-J CO or isotopologues, owing to the difficulty in observing dense tracers.  However, it is more appropriate to analyse the properties of compact clumpy structures as seen in Fig.\,1a using denser gas tracers rather than the diffuse gas traced by low-J CO observation.  Dense gas tracers, such as \cs21\,, HCN\,(1-0) and \co65\, usually trace embedded cloud clumps or cores within a more extended distribution of CO emission \citep{2015ApJ...801...25L, 2011ApJ...735...19S, 2005ApJ...623..826R}, and therefore are very useful for studying the dense, embedded star-forming structures within a much larger molecular region.

In Fig.\,1a, the \co65\ emission peaks are resolved into separate
clumps at S/N = 4, labelled as {\it a}1, {\it a}2, {\it a}3,
{\it a}4, {\it a}5, {\it b}, {\it c} and {\it d}.
For each clump, we list in Table\,2 a number of parameters derived from 
the image in Fig.\,1a for all the clumps except for the clump 
{\it d}.  At the resolution of Fig.\,1a, the clump {\it d} is 
detected only at S/N = 3 and appears to be quite diffuse.  
We therefore derived its parameters from the image in Fig.\,3a, 
which has a coarser resolution of 112 $\times$ 112\,pc.

The size of a clump is specified by its FWHM major and minor 
axes plus the major axis position angle (PA), which are 
respectively given in Columns (2) and (3) of Table\,2. These were 
derived from a 2d Gaussian fit to the clump intensity map.
For the blended clumps ({\it a}4 and {\it a}5), we segmented the cloud into
sub-clouds employing a variant of the CLUMPFIND algorithm
\citep{1994ApJ...428..693W}.
We divided the blended clumps by the half distance of two
peaks and measure their parameters.
We also calculated the effective clump radius $R$, following 
\citet{1987ApJ...319..730S}. After the deconvolution with 
the appropriate ALMA beam (using the CASA function 
``deconvolvefrombeam''), the radii range from 45 to 
180\,pc for the clumps (see Table\,2, Column\,(4)).

We also extracted the 1d spectrum for each clump
within an elliptical aperture with
radii of (major and minor axes (FWHM)) as the oval areas marked in Fig.\,1a.
The resulting spectra are plotted for all the clumps in Fig.\,\ref{spectra}. Using the same elliptical aperture, we derived the integrated CO\,(6-5)
flux from the clump intensity map and 435\um\, continuum
flux density from continuum map.
The line central velocity, the line velocity width
$\Delta V_{\rm FWHM}$, the integrated CO\,(6-5) flux and 435\um\, continuum flux
density given respectively in Columns (5), (6), (8)
and (9) in Table\,2. The resulting $\Delta V_{\rm FWHM}$ ranges
from 60 to 88\,\kms.

We also estimated the Virial and molecular gas masses for each clump.
Following \citet{1981MNRAS.194..809L, 2009ApJ...699.1092H},
the Virial mass ($M_{vir}$) is estimated as
\begin{equation}
	M_{\rm vir}/M_\odot = \frac{5\,(\Delta V_{\rm FWHM}/{\rm \kms})^2\,R/{\rm pc}}{G}, 
\end{equation}
where $R$ is simply the effective clump radius and $\Delta V_{\rm FWHM}$ 
refers to the value of after deconvolution with channel width 13.5\,\kms
and a possible contribution of about 10\kms\ from the disk rotation is further 
removed. (This correction amount was set to the mean velocity 
change over the size of one ALMA beam by examining the P-V plots
of all the clumps. In the Following, the $\Delta V_{\rm FWHM}$ which we have
used in Fig.\,8 and 9 are those corrected ones.)
The derived Virial masses, shown in Column (10)
of Table\,2, rage from $\sim$7 to 60 $\times 10^7\,M_{\odot}$.

We estimated the molecular gas mass of a clump, $M_{\rm mol}$, 
using the 850\um\ continuum flux density-based formula in 
\citet{2016ApJ...820...83S} by converting the observed 435\um\ 
flux density to that at 850\um\, assuming a dust temperature of the 25\,K:

\begin{eqnarray}
  L_{\nu_{850\mum}} &&= 1.19 \!\times\! 10^{27}\!\times\!
  S_{\nu}/{\rm Jy} \!\times\!
	\frac{ (\nu_{850\mum})^{3.8}}{\nu_{obs} (1+z)} 
	\!\times\! \frac{ (d_{L}/{\rm Mpc})^2}{1+z}\nonumber \\
	&&\!\times\! \frac{\Gamma_{\rm RJ} (25,\nu_{850\mum},0)}{\Gamma_{\rm RJ} (25,\nu_{\rm obs},z)}/(\rm erg\, s^{-1}Hz^{-1}) \\\nonumber
\end{eqnarray}
and
\begin{equation}
	\alpha_{\nu}\!\!=\!\!L_{\nu_{850\!\!\mum}}/M_{\rm mol}\!\! =\!\! 6.7\pm 1.7 \!\times\!10^{19} /(\rm erg\, s^{-1}Hz^{-1} M_\odot^{-1}) \\\nonumber
\end{equation}

Where 
{$\Gamma_{\rm RJ} (T_{\rm d},\nu,z)= 
\frac{h\nu/kT_{\rm d}}{e^{h\nu/kT_{\rm d}}-1}$}
and the $S_{\nu}$ is the dust emission flux density.
As the 435\um\ flux density shown in Table\,2 is measured within FWHM (diameter),
it should only account for about 58\% of the total flux. Thus, we
multiply the flux density by 1.731 to estimate the total flux.
For the gas clumps in the nuclear region of NGC\,5135,
$T_d$ could be warmer. In this case, the calculated $M_{\rm mol}$
would be overestimated by roughly a factor of $T_d$/(25\,K).
The resulting $M_{\rm mol}$, given in Column (11) of Table\,2,
ranges from $\sim$1 to 10 $\times 10^8\,M_{\odot}$.

To check the molecular gas mass, we derived the molecular gas mass from the CO flux. We chose the conversion factor $\alpha_{\rm CO} = 0.8$ M$_\odot$(K$\,$km$\,$s$^{-1}\,$pc$^2$)$^{-1}$ \citep{1998ApJ...507..615D}. 
The \co65\,/\co10\ is about 2 (this will be explained in Sec.\,4.3.1).
The molecular gas mass derived this way is consistent with the one derived 
from the dust continuum as shown in Column\,(12) of Table\,2.

Table\,2 shows that all clumps have $M_{\rm mol} > M_{\rm vir}$
except for the clumps {\it b} and {\it d}. The clumps in the former
category (hereafter referred to as Category (i)) are likely to be 
self-gravitationally bound or even undergoing initial collapse. 
On the other hand, the clumps {\it b} and {\it d} in the other 
category (hereafter Category (ii)) would require external pressure 
to remain bound. Interestingly, the Category (ii) clumps are far 
away from on-going SF activity and
also show significantly higher CO\,(6-5)-to-dust continuum flux ratios
(cf. Table\,2, Column (13)) than the clumps in Category (i).

We can also compare the clumps in NGC\,5135 with the molecular gas 
clouds in other galaxies. Fig.\,8 is a plot of the cloud (FWHM) 
velocity dispersion as a function of the cloud radius for the clumps
in NGC\,5135 as well as discrete clouds in the center of the Milky 
Way, nearby spiral galaxies, and two starburst galaxies NGC\,253 and 
IC\,5179.  In comparison to the molecular clouds over the disks of 
nearby normal galaxies, the clouds in the Milky Way center observed by 
\citet{2001ApJ...562..348O} have a larger line width, but a smaller size. In contrast, the clouds in the starburst galaxy NGC\,253 \citep{2015ApJ...801...25L} and the LIRG IC\,5179 \citep{2017ApJ...845...58Z} 
show both larger sizes and broader line widths than clouds
in the Milky Way center, but generally following the lines of equal 
gas surface density for the case of Virialized clouds.  The clouds in
the nuclear region of NGC\,5135 are characterized by still larger sizes and line widths.

Fig.\,9 is a plot of the parameter, $\Delta V_{\rm FWHM}^{2}/R$, 
as a function
of molecular gas mass surface density $\Sigma$ for the same data set as in Fig.\,8. The thick diagonal line shows the locus 
of Virialized clouds. For bound clouds clearly lying above this line,
the cloud velocity width is likely a manifestation of some external
pressure. The dashed curves in Fig.\,9, taken from \citet{2011MNRAS.416..710F},
indicate the relationship between $\Delta V^2_{\rm FWHM}/R$ 
and $\Sigma$ 
for a varying external pressure. 
The Category (i) gas clumps in NGC\,5135
lie around the line tracing the Virial equilibrium. 
In contrast, the two
Category (ii) clumps are clearly located above the Virial equilibrium,
and  require external pressure of the order of 10$^8$ cm$^{-3}$K in 
order to remain bound.

\subsection{CO\,(6-5) Emission \label{CtD}}

\subsubsection{CO\,(6-5) Emission to Continuum Ratio\label{sect4.3.1}}

On galaxy scale, the ratio of the \co65\ line luminosity, 
$L_{\rm CO(6-5)}$, to $L_{\rm IR}$ varies only by up to 30\% among
local LIRGs and shows little dependence on $L_{\rm IR}$ or 
the FIR color \citep{2014ApJ...787L..23L, 2017ApJS..230....1L}.
This strongly 
requires that the energy sources for both the \co65\ and 
the dust emissions are ultimately tied to the same SF process. 
This narrows down the candidate heating mechanisms for the \co65\
emission to fewer choices, including photon heating in 
the photon dominant regions (PDRs) around young massive stars
and SN-powered shock heating.

Our recent ALMA observation of nearby LIRGs show that, on 
scales of 100\,pc or less, local peaks of the \co65\ emission
do not always have corresponding peaks of the 435\um\ dust 
continuum emission.  Such examples include NGC\,7130
 \citep{2016ApJ...820..118Z}, IC\,5179 \citep{2017ApJ...845...58Z}
and the case of NGC\,5135 shown here. Under the assumption 
of a constant dust-to-gas abundance ratio, the spatial peaks 
of the two emissions should follow each other if both the dust 
and \co65\ emissions are related to the same photon heating.
This finding therefore favors the SNe-powered shock 
gas heating scenario for the \co65\ emission.
Here we investigate further this subject in the case of NGC\,5135.

As shown in Column (13) of Table\,2, the \co65\ flux to 
the 435\um\ continuum flux density ratio, $R_{\rm CO/cont}$,  
varies among the CO\,(6-5) clumps. Further more, while 
the Category (i) clumps satisfy $600 \lesssim R_{\rm CO/cont}
\lesssim 1800$\,\kms, the two Category (ii) clumps have 
$R_{\rm CO/cont} > 4,000$\,\kms. It is evident in Table\,2,
the higher $R_{\rm CO/cont}$ values associated with 
the Category (ii) clumps are mostly due to the unusually 
faint dust continuum emission at 435\um. One can express 
\begin{equation}
  R_{\rm CO/cont} \propto (f_{\rm CO(6-5)}/f_{\rm CO(1-0)})\,(M_{\rm gas}/M_{\rm dust})\,T^{-1}_{\rm d}, 
\end{equation}
where we have assumed the CO\,(1-0) flux, $f_{\rm CO(1-0)}$, scales 
with the molecular gas mass $M_{\rm gas}$. This shows that a 
higher $R_{\rm CO/cont}$ can stem from either a hotter CO gas 
or/and a cooler dust temperature. In the nuclear region of NGC\,5135,
the variation of $T_d$ is limited to, perhaps, a factor of 3
(i.e., from 15 to 50\,K) at most. The observed variation of 
$R_{\rm CO/dust}$ is a factor of $\sim$5 in Table\,2, mainly 
between the two clump categories. A comparable variation 
is also seen in the case of IC\,5179 (Zhao et al. 2017). Therefore,
it requires a modest variation of a factor of 2 or so in 
$f_{\rm CO(6-5)}/f_{\rm CO(1-0)}$ in order to explain the observation.
If the \co65\ emission is associated with SNe-shock heating, 
the ideal location for a higher $R_{\rm CO/cont}$ ratio is where
massive O star formation has ended while SNe activity is still 
strong, a scenario we discussed in the case of NGC\,7130 (Zhao et 
al. 2016).  A necessary condition for the validity of this scenario 
is that some dense gas can survive the massive star formation,
which might be possible in dense and clumpy ISM.

\subsubsection{\co65\ Emission and Current SF\label{sect4.3.2}}

In Fig.\,\ref{align1}, the black contours of the integrated \co65\ 
line emission are overlaid on a ground-based 8.7\um\ image 
\citep{2008ApJ...685..211D} on the left side, and on an 
{\it HST} Pa-$\alpha$ image \citep{2006ApJ...650..835A} on the right.
In both plots, we also show the low-resolution Chandra 0.4-8 keV
broadband X-ray emission (Levenson et al. 2004) in red 
contours.  This X-ray emission is mostly associated with a hot, 
ionized gas powered by SNRs (Levenson et al 2004; 
Colina et al. 2012).

The 8.7\um\ image is dominated by the mid-IR emission bands from
the so-called Polycyclic Aromatic Hydrocarbons (PAHs). 
The Pa-$\alpha$ emission is caused by the ionizing UV radiation from massive O stars and is regarded as a reasonable tracer of the on-going SF activity (timescale $\sim$ 10\,Myr). 
The PAH emission traces SF of 10 times longer time scale 
($\sim$100\,Myr), as PAH molecules are mostly heated by non-ionizing B stars 
\citep{2008ApJ...685..211D}. As shown in Fig.\,\ref{align1}, 
the PAH and Pa-$\alpha$ emissions show similar surface brightness
distributions. In contrast, the overall spatial morphology of 
the \co65\ emission appears to be different from that of either
the PAH or the Pa-$\alpha$ emission. However, the Category (i) 
\co65\ clumps are all relatively close to local emission peaks 
of the PAH or Pa-$\alpha$ emission whereas the Category (ii) 
\co65\ clumps are significantly farther away from any bright 
PAH or Pa-$\alpha$ peak.  Therefore, it is reasonable to expect 
a much weaker far-UV radiation intensity at the location of each
Category (ii) clump. This naturally explains why the dust emission
is unusually faint at each of the Category (ii) clumps.

\subsection{Possible Heating Scenarios for \co65\ Emission \label{PtH}}
\subsubsection{SN-powered Shock Heating Scenario\label{sect4.4.1}}

With an integral field spectrograph, \citet{2012ApJ...749..116C} 
measured the intensity and velocity fields of both the Br-$\gamma$
and the \FeII\ 1.64\um\ emission lines in the nuclear region of 
NGC\,5135. While the Br-$\gamma$ traces the current star formation, 
the \FeII\ line emission is regarded as a particularly good 
tracer of SNRs \citep{1991ApJ...383..164G}. 
In Fig.\,11, we show a plot of the \FeII-to-Br-$\gamma$ line
ratio versus the CO\,(6-5)-to-continuum flux density ratio 
for all the clumps listed in Table\,2, except for clump 
{\it d}, which is located outside the field of view of 
the \FeII\ observation. We also indicated 
the typical \FeII-to-Br-$\gamma$ line ratios for different
astrophysical objects, taken from \citet{2014MNRAS.438..329F}. 
Note that the line ratio range shown for Seferts 
is largely irrelevant here as our molecular clouds are  
all located far away from the AGN.

A number of studies have attempted to identify the physical 
causes behind the observed variations of the \FeII-to-Br-$\gamma$ 
line ratio (e.g., \citealt{1997ApJ...482..747A, 1988A&A...203..278M,
1990ApJ...360...55M, 1993ApJ...406...52M, 1991ApJ...383..164G, 
2004A&A...425..457R, 2005MNRAS.364.1041R, 2006ApJ...645..148R, 
2009ApJ...694.1379R, 2013MNRAS.430.2002R, 2014MNRAS.438..329F}). 
From these studies,
two main conditions for an enhanced \FeII\ emission relative
to a Hydrogen recommbination line emission emerge: (a) Presence of 
shocked gas and (b) favorable environment for the gas-phase Fe 
to be abudant in the form of Fe$^{\rm +}$. Iron is normally depleted 
onto grains in the interstellar gas phase, so fast shocks, such 
as those associated with SNe, can cause grain destructions and 
therefore enrich gas-phase Fe abundance. Another important 
prerequisite for a strong \FeII\ line is an ionization field 
in favor of Fe$^{\rm +}$. 
Given the low ionization potential of Fe$^{\rm +}$
(16.2 eV), most of Fe is in higher ionization states in HII
regions. In comparison, partially ionized gas 
in SNRs and Fe$^{\rm +}$ is believed to be abundant 
\citep{1988A&A...203..278M}.
The collisional exitation with electrons could therefore make 
the \FeII\ line much brighter in SNRs.

The \FeII-to-Br-$\gamma$ line ratios for the Category (i) clumps
in NGC\,5135 are around 3, which is just outside 
the upper tip of the range for Galactic H{\scriptsize{II}} regions.  
These ratios are slightly higher than those seen in the nuclear 
star-forming regions in the Seyfert galaxy NGC\,613 \citep{2014MNRAS.438..329F}, suggesting some mild enhancement in the \FeII\ 
line emission for the gas clouds in NGC\,5135. In comparison,
this line ratio for the Category (ii) clump {\it b} equals 
6.5, implying a factor of 2 further enhancement in the relative
\FeII\ emission from the Category (i) clouds. 
It is not surprising for the observed line ratio of 
the Category (ii) cloud to be smaller than 
the typical values seen in Galactic SNRs because we are averaging 
over a much larger area than the typical size of Galactic SNRs
and also because there is still low-surface brightness star 
forming activity near the cloud (see Fig.\,10). The observed trend
 in Fig.\,11 indicates that the same SNe shocks are 
likely to play a positive role in the observed variations in both 
 \FeII/Br-$\gamma$ and CO\,(6-5)/dust ratios.

Additional evidence in favor of the SNe-shock heating scenario for
the \co65\ emission in NGC\,5135 includes (a) the prevailing X-ray 
emission from the hot, ionized gas excited via SN shocks (Colina 
et al. 2012); and (b) that there is a very good velocity field correspondence
between the \co65\ clumps and that of the underlying \FeII\ 
emission: the nominal \co65\ and \FeII\ line velocity offset
varies around the mean of -142\kms\ by only a few \kms\ among the clumps.
(Note that the mean velocity field difference is likely a result of 
the different velocity reference frames adopted.)
This velocity correspondence suggests that the warm CO gas and 
the shocked/ionized gas are reasonably well mixed with each other 
in space and velocity field. Another independent evidence in favor 
of the SNe-shock 
heating scenario is the global tight correlation between the IR dust 
emission and the mid-$J$ CO line emission shown by Lu et al. (2017), 
which requires that the gas heating ultimately derives the energy 
from the same SF process. The SNe heating scenario would naturally 
fit this requirement.

\subsubsection{Bar-Induced Shock Heating Scenario \label{sect4.4.2}}

Fig.\,\ref{align} displays the integrated \co65\ line emission contours
overlaid on the {\it HST} F606W (0.606\um) image \citep{1998ApJS..117...25M} 
and {\it HST} F160W (1.60\um) image \citep{2006ApJ...650..835A}, 
respectively.  The {\it HST} images aligned with our CO\,(6-5) data by matching 
our adopted AGN position with the brightest point in each optical image.
As already mentioned before, the \co65\ emission has a good
spatial correspondence 
with the dust lanes that can be seen in the optical and near-IR continuum
images here. These roughly symmetrical dust lanes are induced 
by the inner stellar bar, both of which are more visible in a larger 
UV/optical image such as the one shown by \citet{1997ApJ...482L.135M}.
This correspondence between the \co65\ emission and the bar-induced 
dust lanes in NGC\,5135 is similar to that observed in NGC\,7130, another 
LIRG with a strong stellar bar \citep{2016ApJ...820..118Z}.  
Indeed NGC\,5135 and NGC\,7130 have many similarities:  both are LIRGs 
with a strong circumnuclear star formation and a Seyfert\,2 nucleus.
Circumnuclear dust lanes have been found in many spiral galaxies, though 
strong 2-arm dust lanes are found only in barred galaxies such as 
NGC\,5135 and NGC\,7130 \citep{2003ApJS..146..353M}.

According to the model of \citet{1992MNRAS.259..345A}, the 2-arm
dust lanes are associated with the shock fronts triggered by the presence 
of a bar in a rotating gas disk. Thus, bar-induced shocks could be
possible in the nuclear region of NGC\,5135.
Inside the dust lanes, the gas (and dust) density is significantly enhanced, 
but SF is suppressed by strong shears \citep{1992MNRAS.259..345A}.
This seems to imply that the warm dense gas traced by \co65\, and 
the SF regions traced by Pa-$\alpha$ are not related to each other.

In order to account for the tight correlation between the \co65\ 
emission and the total dust emission on galaxy scale for LIRGs, one 
has to relate the \co65\ emitting gas to the SF activity in this 
scenario.  It is still possible that the warm dense gas and the SF 
regions are related to each other, albeit their positions are 
slightly offset. It is known that some galaxies with weak stellar 
bars (therefore weaker shears in dust lanes) have SF in their dust lanes
\citep{1982A&A...114....7C, 2003ApJS..146..353M}.
Hypothetically, one can envisage the following scenario: First, SF 
does occur in clouds of dense gas formed in the post-shock gas down 
stream from the bar-induced shock front (the dust lane). Then these
dense gas clouds will be rapidly consumed/destroyed by the SF and 
the associated feedback.
In this scenario, under the assumption that the destruction time scale
of the dense clouds is much shorter than the SF time scale associated 
with the Pa-$\alpha$ emission (a few Myrs),  the spatial offset is 
the product of the SF time scale times the downstream velocity of 
the post-shock gas, which is a few 10s\,\kms \citep{1992MNRAS.259..345A}.
This indeed results in an estimate for the offset of $\lesssim$100\,pc.
It is worth noting that similar offset between H{\scriptsize{II}} 
regions and dust lanes associated with
spiral arms in grand-design galaxies such as M\,51 have been found
in the literature, and \citet{2001AJ....122.3017S}
argued that it implies that the H{\scriptsize{II}}
regions develop subsequent to the time of maximum concentration
of the dust and molecular clouds. 

However, it is unclear how this bar-induced shock heating scenario
for the CO\,(6-5) emission can be made to explain the similar 
variation in the CO\,(6-5)-to-continuum flux ratio seen in IC\,5179
(Zhao et al. 2017), which does not have a strong stellar bar.

\section{SUMMARY \label{sum}}

In this paper we present the results from our ALMA observations 
of the \co65\ line and its underlying dust continuum at 435\um\
in the nuclear region of the nearby LIRG, Seyfert\,2 galaxy 
NGC\,5135, at a physical resolution of 48 $\times$ 40\,pc.  
Our main findings are:

(1) The central AGN is undetected in either the 435 $\um$ dust 
continuum or \co65\ line emission if its line 
velocity width is no less than $\sim$40\,\kms, 
resulting in an AGN contributes at most 1\% of the integrated circum-nuclear
\co65\ flux seen in our ALMA observation. On the other hand,
the non-detection in continuum emission may simply reflect the lack of 
sensitivity in our observation. 

(2) The circum-nuclear \co65\ emission is resolved into gas 
clumps of radii of 45-180\,pc and line velocity widths of 
60-88\,\kms. While the clump sizes are only slightly larger 
than typical giant molecular clouds in nearby spiral galaxies,
their velocity widths are significantly higher.  They fall 
into two categories: (i) The 5 clumps that are near some current
star formation activity are likely to be in Virial equilibrium,
and (ii) the other 2 clumps without clear current star formation
activity nearby seem to be unbound unless there is significant 
external pressure. 

(3) The clumps in Category (ii) have much higher \co65\ to 
dust continuum ratios than those in Category (i). Furthermore,
the CO\,(6-5)-to-continuum ratios show an increasing trend with 
the \FeII-to-Br-$\gamma$ ratios, which 
we interpret as evidence for supernova-driven shocked gas providing a significant contribution to the \co65\ emission. 

(4) The clumps are distributed along the symmetric optical dust
lanes associated with the stellar bar at the center of the galaxy.
Like NGC\,7130, another barred Seyfert galaxy, the gas concentrations
could be a result of the bar-induced instability and is subject 
to bar-induced shock heating.

\section*{Acknowledgments}

We thank an anonymous referee for a number of very constructive comments.
We thank Drs. Cheng Cheng, Luis Colina, Adam Leroy,
Claudio Ricci, and Chentao Yang for their insightful 
comments and/or useful communications during the preparation
of the manuscript.
This paper makes use of the following ALMA data: 
ADS/JAO.ALMA\#2013.1.00524.S. ALMA is a partnership of ESO  
(representing its member states), NSF  (USA), and NINS  
(Japan), together with NRC  (Canada) and NSC and ASIAA  
(Taiwan), in cooperation with the Republic of Chile. 
The Joint ALMA Observatory is operated by ESO, AUI/NRAO, 
and NAOJ. This work is supported in part by the National Key R\&D Program of 
China grant \#2017YFA0402704, the NSFC grant \#11673028 and \#11673057,
and by the Chinese Academy  of Sciences  (CAS),
through a grant to the CAS South America
Center for Astronomy  (CASSACA) in Santiago, Chile.
C.C. acknowledges support by NSFC grant No.\,11503013.  
Y.G. acknowledges support by NSFC grants No.\,11173059, 11390373, and 11420101002. 
H.W. acknowledges support by NSFC grant No.\,11733006.
V.K. acknowledges support from the FONDECYT grant No.\,3160117.
T.D.-S. acknowledges support from ALMA-CONICYT project 31130005 and FONDECYT regular project 1151239.

\bibliographystyle{apj}
\bibstyle{thesisstyle}
\bibliography{main.bib}

\newpage

\begin{figure*}[h!tb]
\centering
\includegraphics[width=6.9in]{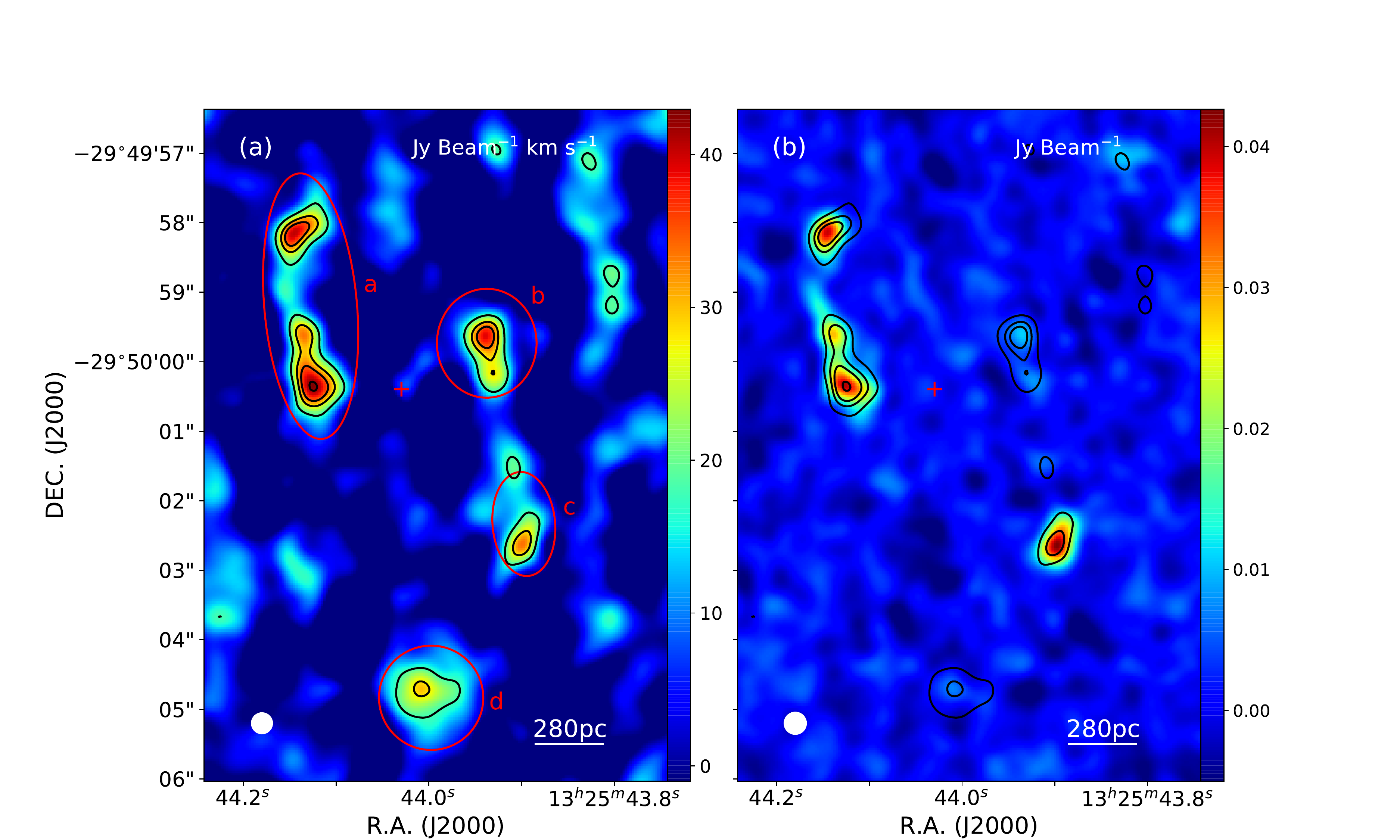}
\caption{Same as the panels (a) and (b) in Fig.\,1, respectively, but using 
the CO\,(6-5) and continuum data at a larger effective beam.
The contour levels are [3, 4, 5, 6]$\times \sigma$ 
($\sigma = 8$\,Jy\,beam$^{-1}$\kms,  with the beam size of 
0.4\arcsec\ $\times$ 0.4\arcsec\ here as shown by the filled ellipse
in white in each panel. 
\label{figure2}}
\end{figure*}

\begin{figure*}[!htbp]
\begin{center}
\includegraphics[width=6.5in]{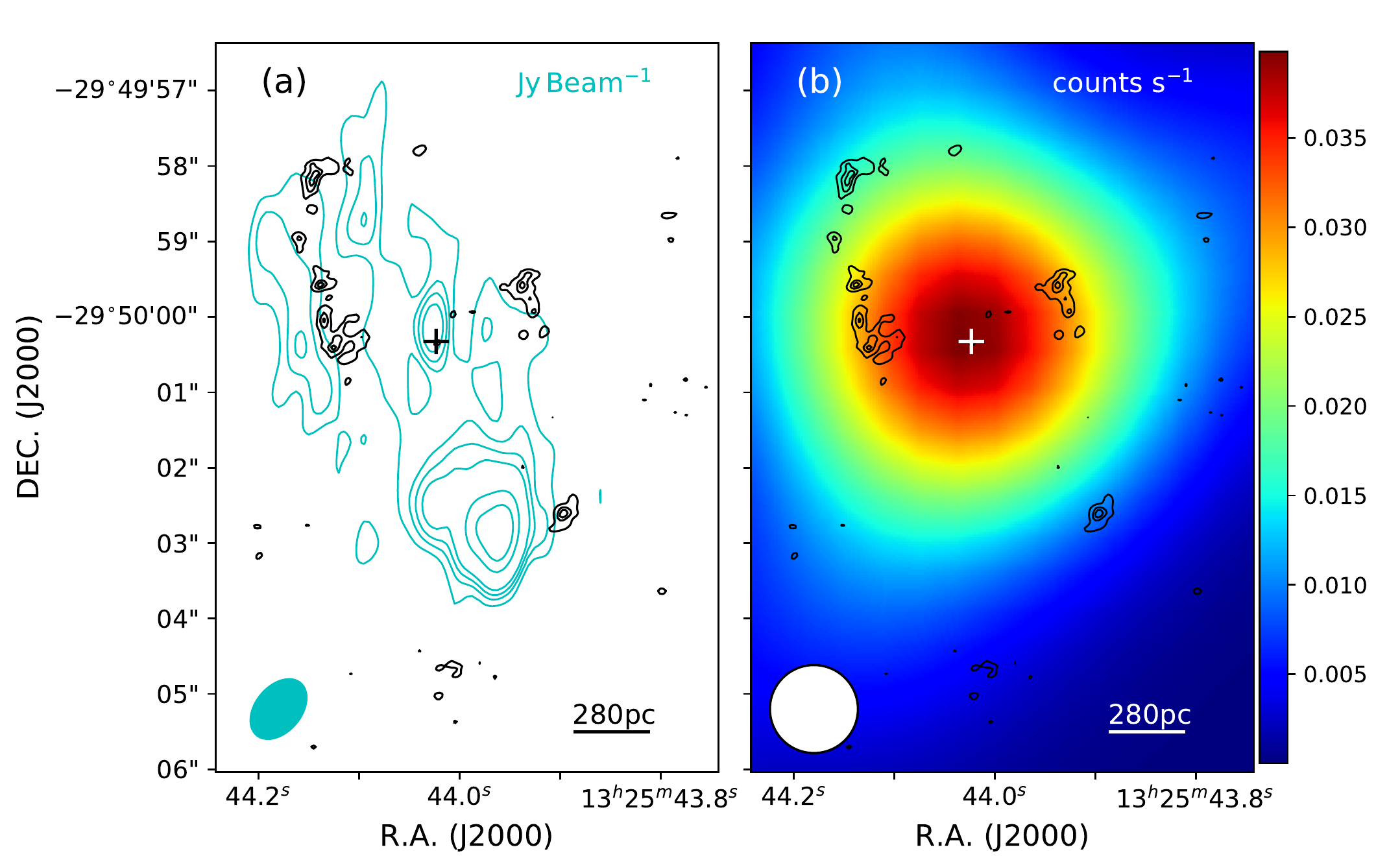}
\caption{Black contours of the integrated CO\,(6-5) line emission overlaid on 
(a) an VLA radio
cyan contours at 6cm (the cyan contour levels are [3, 5, 7, 10, 30, 50]$\times \sigma$ ($\sigma$ =1.2e-04$\,$Jybeam$^{-1}$) and (b) a Chandra 4-8\,kev X-ray image. 
The black contour level are [3, 5, 6]$\times \sigma$ ($\sigma$=1.2$\,$Jybeam$^{-1}$kms$^{-1}$). The white plus sign in each panel presents the adopted AGN location.
\label{figure3}}
\end{center}
\end{figure*}

\begin{figure*}[!htbp]
\begin{center}
\includegraphics[width=6.9in]{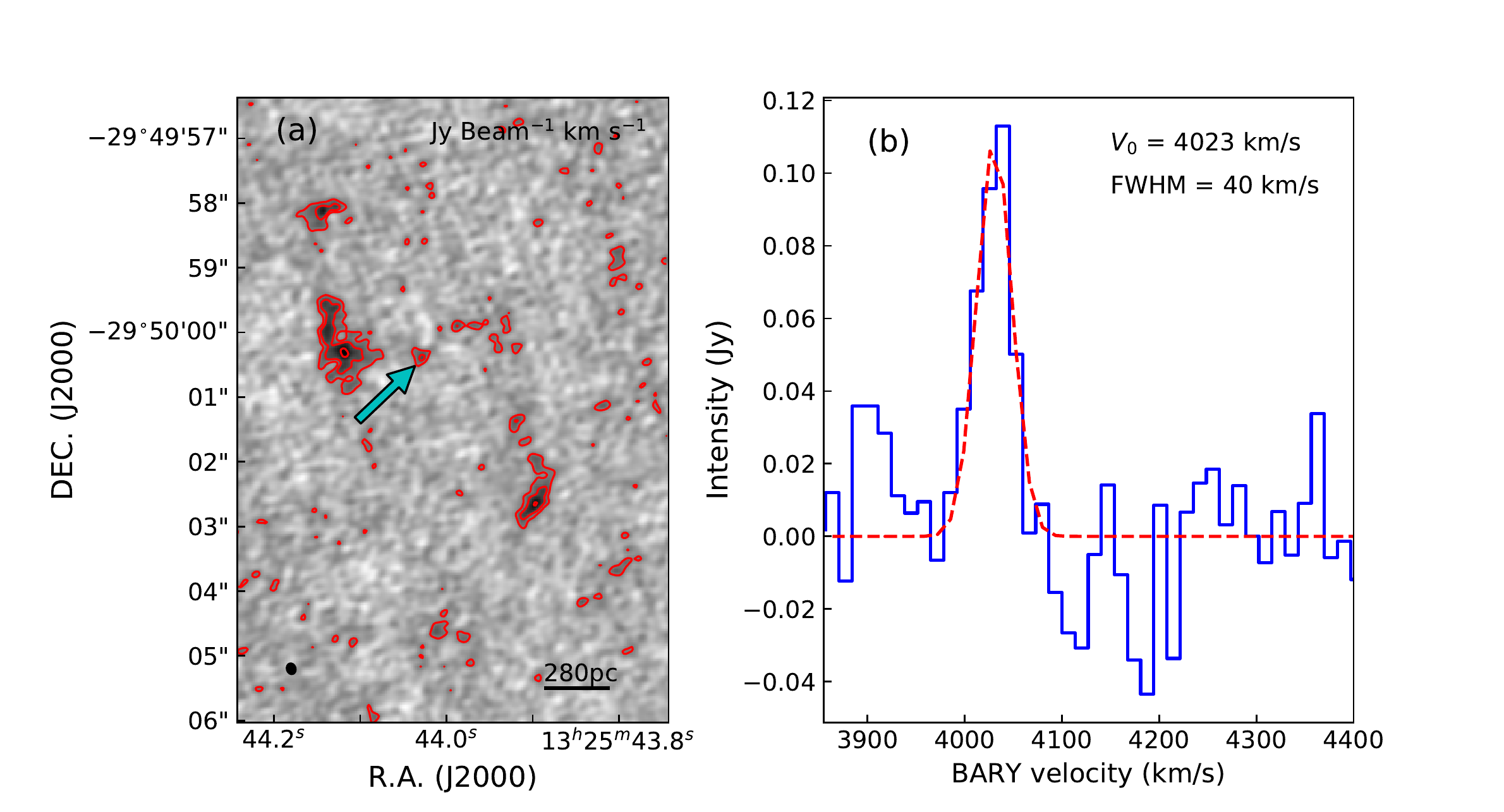}
\caption{
(a) Contours of the integrated CO\,(6-5) from 4019 to 4032 km/s on the image of the same integrated CO\,(6-5) intensity. The cyan arrow points to the central AGN. The contour levels are [3, 5, 8]$\times \sigma$ ($\sigma$=0.5$\,$Jybeam$^{-1}$kms$^{-1}$). And (b) the CO\,(6-5) spectrum at the central AGN. The central velocity ($V_0$) and FWHM of a Gaussian fit are given in the plot.   
\label{agnspec}}
\end{center}
\end{figure*}

\begin{figure*}[!htbp]
\begin{center}
\includegraphics[width=6.9in]{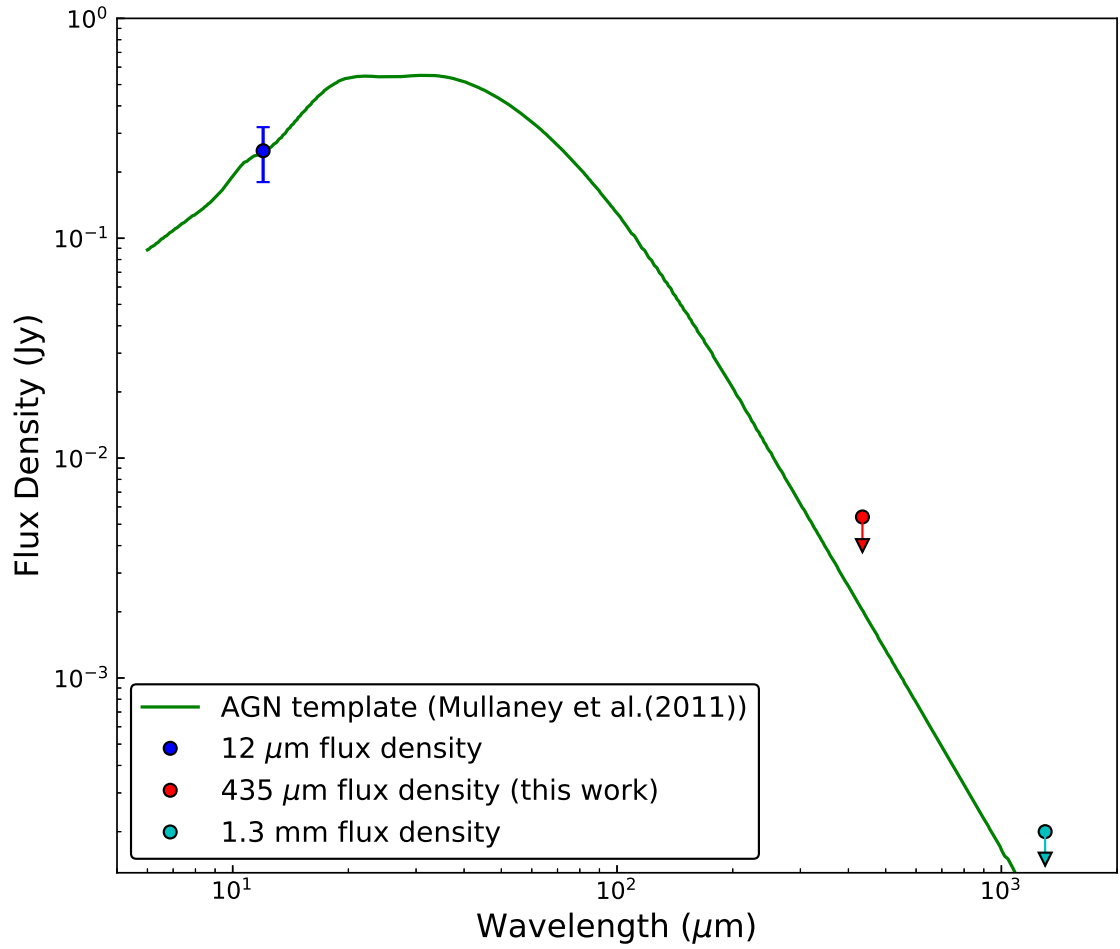}
\caption{
Plot of the empirical infrared spectrum (the green curve) of the AGN in NGC\,5135,
which is based the observed X-ray luminosity and anchored at the 12\um\ 
flux density measurement (the filled circle in blue).  Also shown are two 
ALMA flux upper limits at 435\um\ and 1.3\,mm, respectively (see the text).
\label{agntem}}
\end{center}
\end{figure*}

\begin{figure*}[!htbp]
\centering
\includegraphics[width=6.9in]{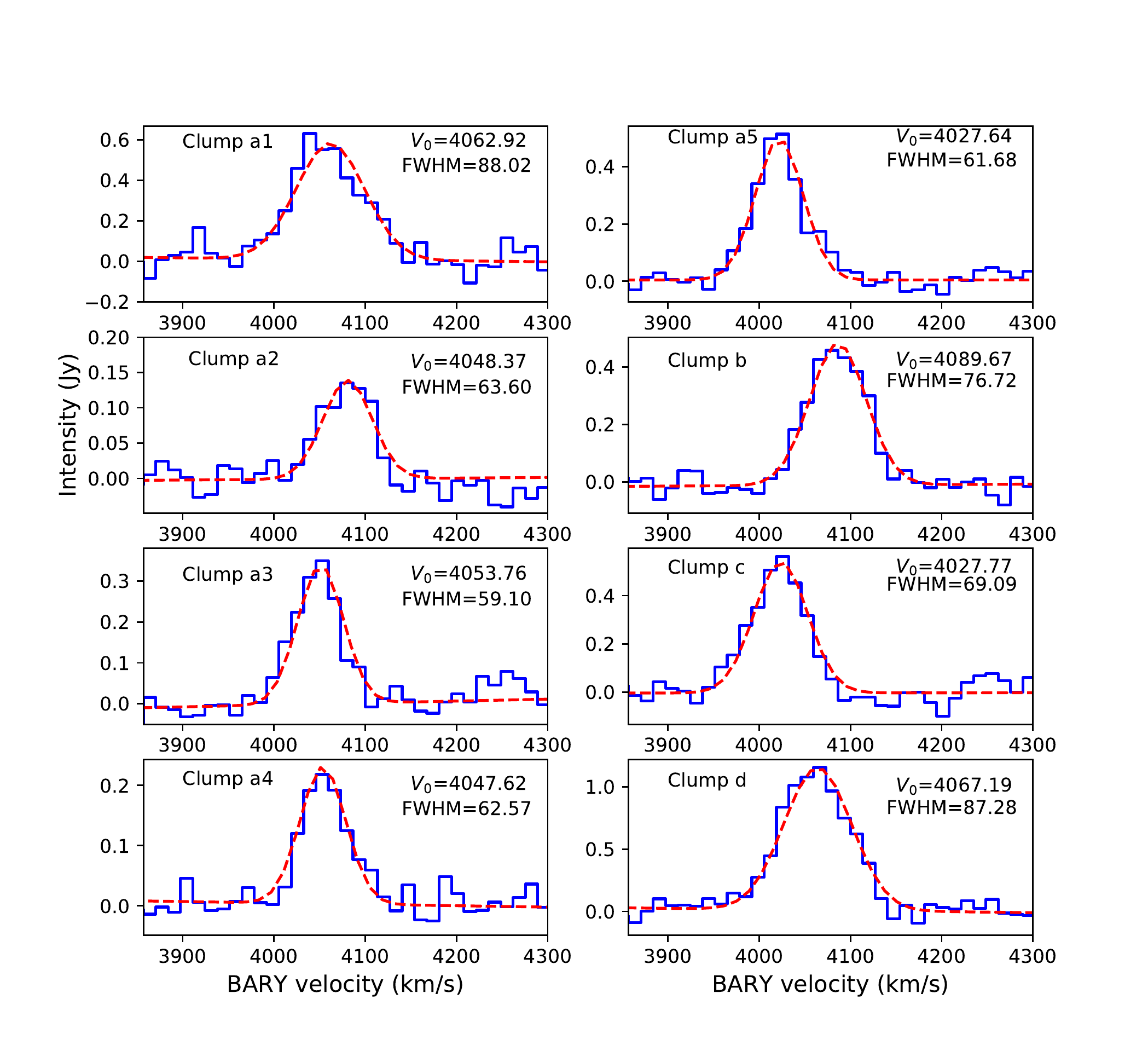}
\caption{ Spatially integrated CO\,(6-5) line profile of
  various clumps (a1, a2, a3, a4, a5, b, c, d). The central velocity ($V_0$)
  and FWHM of a Gaussian fit are given in each plot.
\label{spectra}}
\end{figure*}

\begin{figure*}[!htbp]
\centering
\includegraphics[width=6in]{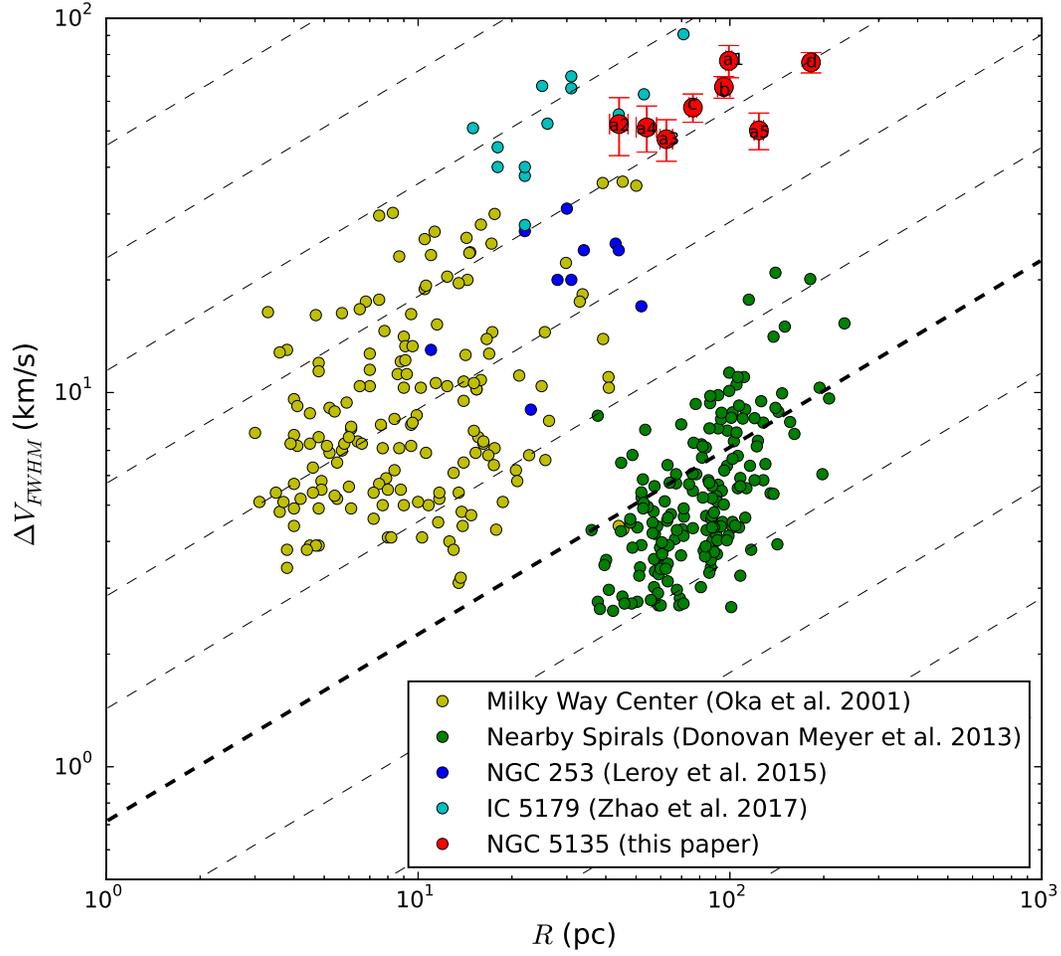}
\caption{Plot of the CO\,(6-5) line width, $\Delta V_{\rm FWHM}$, as a function of the cloud 
radius $R$ for a sample of giant molecular clouds and gas complexes in 
the Milky Way and various local galaxies, adopted from Leroy et al. (2015).
As a comparison, our NGC\,5135 clumps are added (i.e., large filled circles
labeled by the clump number).
The light dashed lines follow $\Delta V_{\rm FWHM} \propto$  R$^{0.5}$, the relation
expected for Virialized clouds with a fixed surface density $\Sigma$,  spaced 
by a factor of 2 vertically.  The thick dashed line is for  $\Sigma$ $\approx$ 
285 M$\odot$ pc$^{-2}$.
\label{dis_size}}
\end{figure*}

\begin{figure*}[!htbp]
\centering
\includegraphics[width=6in]{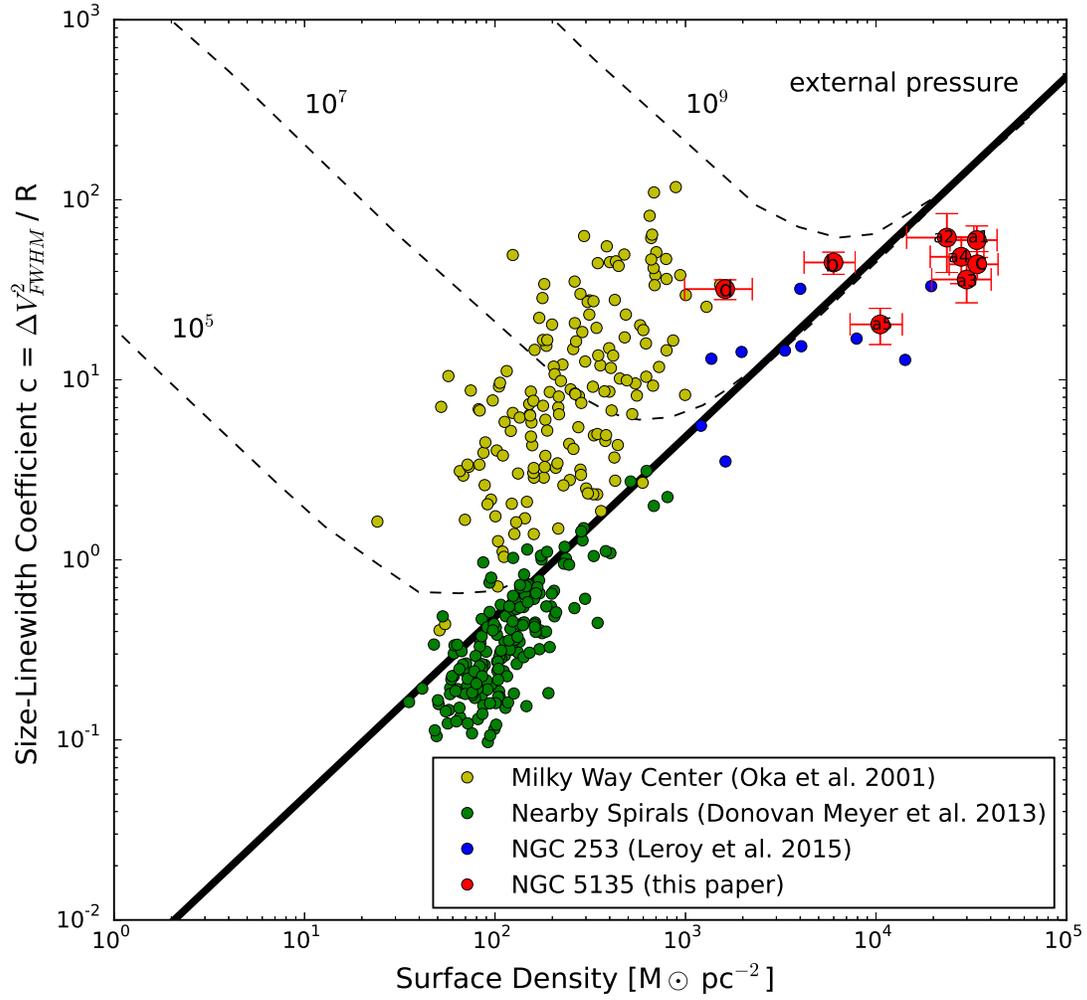}
\caption{
Plot of the cloud radius-line 
width coefficient, $\Delta V^2_{\rm FWHM}/R$, as a function of the gas surface density
$\Sigma$, also adopted from Leroy et al. (2015), for the same data set 
as in Fig.\,7.  The gas clouds that are 
in Virial equilibrium (i.e., $M_{\rm mol} = M_{\rm vir}$) follow the thick
line. Bound clouds with $M_{\rm mol} < M_{\rm vir}$ follow one of the thin
curves representing various external pressures as labelled in terms of 
P/k$_B$ (in units of cm$^{-3}$ K). 
\label{sc}}
\end{figure*}

\begin{figure*}[!htbp]
\begin{center}
\includegraphics[width=7.0in]{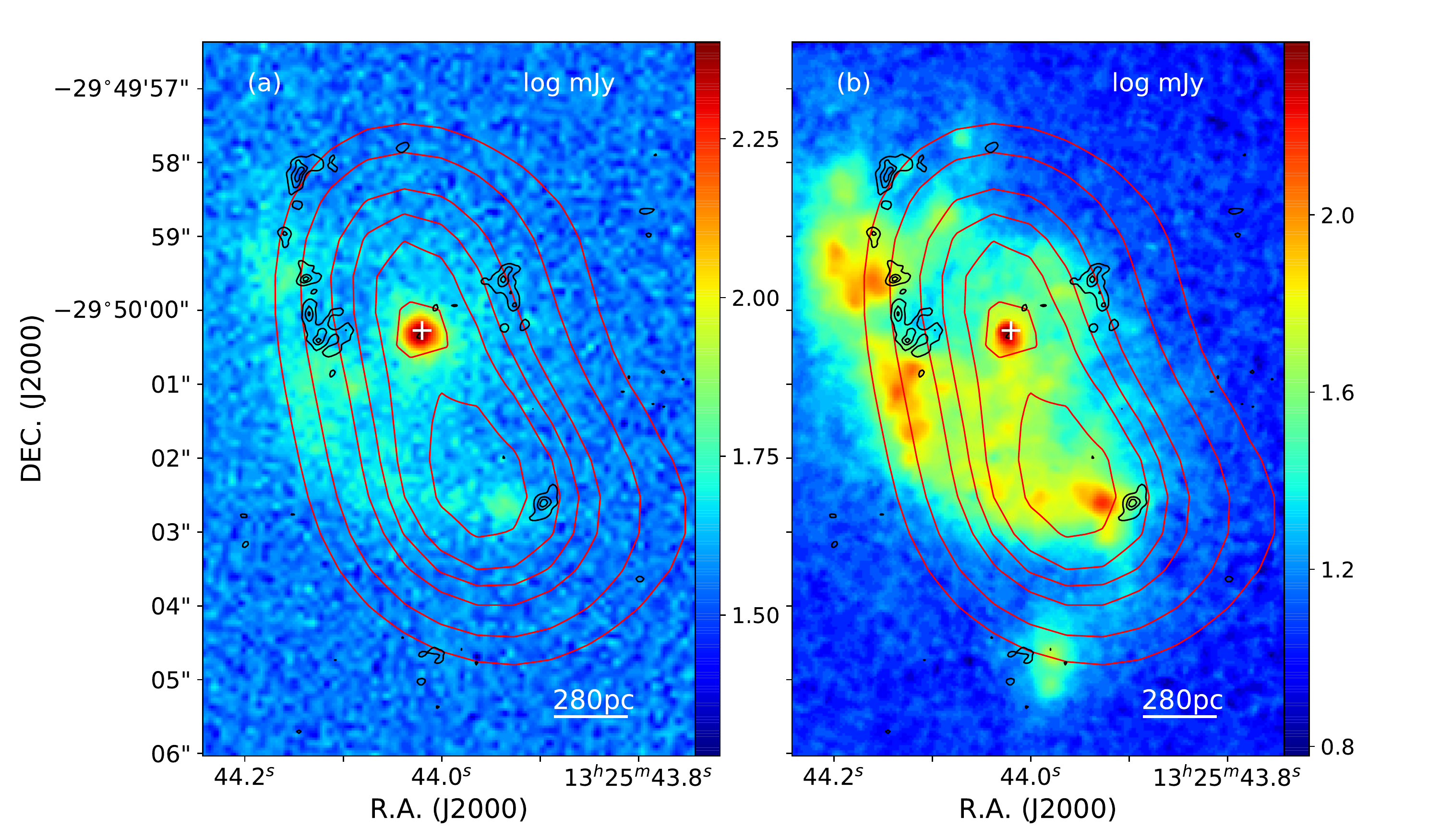}
\caption{
Integrated CO\,(6-5) line emission contours, at [3, 5, 6]$\times \sigma$  
(where $\sigma$=1.2$\,$Jybeam$^{-1}$kms$^{-1}$),  overlaid on (a) 
a 8.7\um\ image dominated the PAH emission (in log scale) and
(b) an image of the Pa-$\alpha$ line emission (in log scale).
The red contours in each panel stands 
for the X-ray intensity of NGC\,5135 (with the contours at
5, 20, 30, 40, 70, 150, 260 counts), obtained in the Chandra 0.4-8$\,$keV 
band by \citet{2004ApJ...602..135L}. The white plus sign in each panel 
marks the adopted AGN location.
\label{align1}}
\end{center}
\end{figure*}

\begin{figure*}[!htbp]
\begin{center}
\includegraphics[width=6.0in]{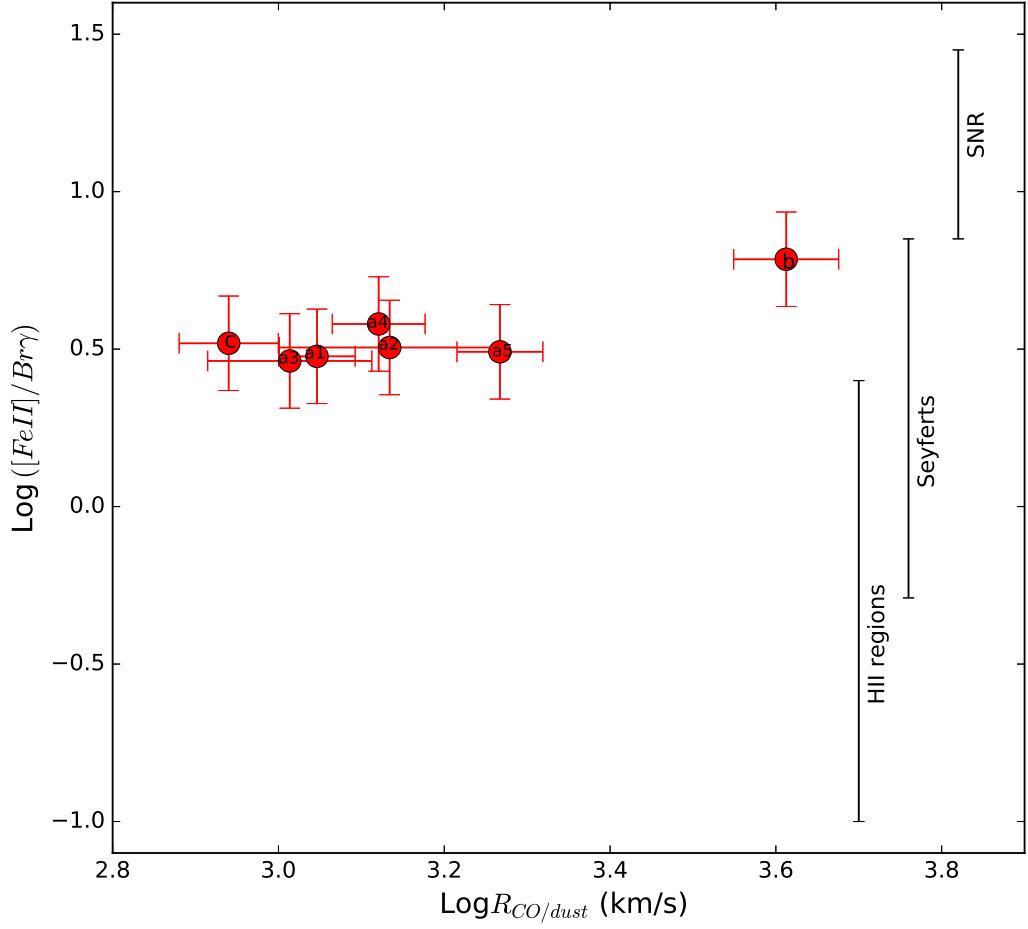}
\caption{
Plot of the \FeII\ 1.46\um-to-Br-$\gamma$ line ratio as a function of
the CO\,(6-5)-to-continuum flux ratio for the NGC\,5135 clumps as labelled.
The extinction corrected \FeII\ and Br-$\gamma$ line surface brightnesses
at the local of a clump are estimated from \citet{2009ApJ...698.1852B} and 
the typical error is 0.15\,dex.
The typical ratios for different types of astrophysical
objects are noted in the plot (see the text).
The clump {\it d} is not plotted here as it does not 
have the corresponding \FeII\ or Br-$\gamma$ data.
\label{ratio}}
\end{center}
\end{figure*}

\begin{figure*}[!htbp]
\begin{center}
\includegraphics[width=6.9in]{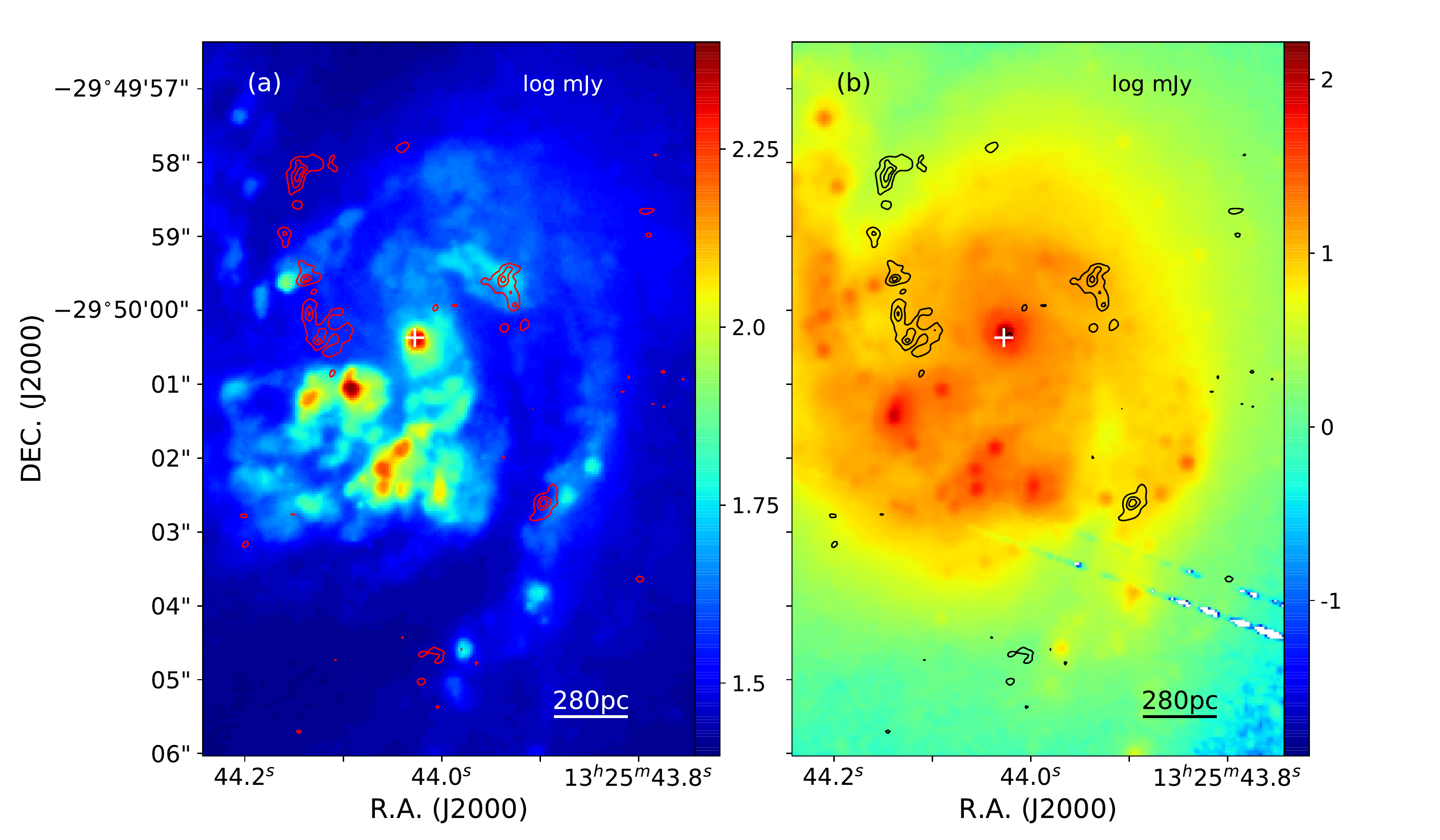}
\caption{
Integrated CO\,(6-5) line emission contours, at [3, 5, 6]$\times 
\sigma$ (where $\sigma$=1.2$\,$Jybeam$^{-1}$kms$^{-1}$), overlaid on 
(a) an {\it HST} F606W (0.606\um, in log scale) image and
(b) an {\it HST} F160W (1.60\um, in log scale) 
image. The white plus sign in each panel marks the AGN position.
         \label{align}}
\end{center}
\end{figure*}

\newpage

\begin{table*}
\centering
\caption{Basic properties of NGC 5135 and ALMA Observation Log}
\label{tab:example_table}
\begin{tabular}{lccccccc} 
\hline\hline 
 &  & &  Basic Properties  &  &   &   & \\
\hline
Name & R.A. (J2000) & Dec. (J2000)&Dist. & cz
& Morph & Spectral Type & $\log\,L_{\rm IR}$ \\
 & (hh:mm:ss)& (dd:mm:ss) &Mpc &kms$^{-1}$ & & & $L_{\odot}$\\
 (1)& (2)& (3)&  (4) & (5) & (6)& (7) & (8)\\
\hline
NGC 5135 & 13:25:43.99 &-29:50:01.06 & 59 & 4105
& SB (s)ab & Sy 2 & 11.33 \\ \hline
\hline\hline
&  & &  ALMA observation log  &  &   &   & \\
\hline
SB & Date & Time (UTC) & Configuration &N$_{ant}$
& l$_{max}$ & t$_{int}$ & T$_{sys}$\\
  & (yyyy/mm/dd)& &  &  &   (m) &  (seconds) & (K)\\
(1)& (2)& (3)& (4)& (5)& (6)& (7)& (8)\\
\hline
Xa216e2$\_$Xcb0 & 2015/06/02-2015/06/03 & 23:58:10-00:51:44 & C34-5
& 39   & 885.6  &   21.03    &  935 - 839   \\
\hline
\end{tabular}\\
\tablecomments{In the upper table section on galaxy basic properties:  Col. 1: source name;
Cols. 2 and 3: right ascension and declination;
Col. 4: distance; Col. 5: heliocentric velocity; Col. 6: galaxy optical morphology; 
Col. 7: Nuclear activity classification; Col. 8: the total infrared luminosity (8-1000$\,\mu$m). \ \ \ 
In the lower table section on ALMA observation log: Col.\,1: schedule-block number;
Cols.\,2 and 3: observational date and time; Col.\,4: observational configuration;
Col.\,5: number of usable 12-m antennae  (i,e,. un-flagged);
Col.\,6: maximum baseline length; Col.\,7: on-source integration time;
Col.\,8: median system temperature.}
\end{table*}

\begin{table*}
\centering
\caption{ Physical properties of the individual clumps in our CO\,(6-5) image}
\label{tab:example_table}
\begin{tabular}{lcccccccccccc} 
\hline\hline %
No.& Size & PA & Radius & $V_0$ & $\Delta V_{\rm FWHM}$ & f$_{peak}$ & f$_{\rm CO(6-5)}$ & I$_{\rm cont}$ & M$_{\rm vir}$ & M$_{\rm mol}$ & M$^*_{\rm mol}$ &R$_{\rm CO/cont}$ \\
 & $''$$\times$$''$ & Deg & pc & kms$^{-1}$ & kms$^{-1}$ & Jy\,beam$^{-1}$  & Jy$\,$kms$^{-1}$
& mJy & $\times$10$^7$\,M$\odot$ &$\times$10$^8\,$M$\odot$ & $\times$10$^8\,$M$\odot$ & kms$^{-1}$\\
&  &  &  &  &  & \kms &  &  &  &  &  &  \\
(1)& (2)& (3)& (4)& (5)& (6)& (7)& (8)&(9)&(10)&(11)&(12) &(13)\\ \hline
 a1 &  0.65$\times$0.34 & 16.50  & 99.2   & 4062.9   & 88.0  & 13.3 & 91.1  & 81.8   & 37.0   & 10.5 &12.6 & 1113.0\\
  & $\pm$0.019$\times$0.010 & & $\pm$2.2  & $\pm$3.0 & $\pm$7.6 &    & $\pm$3.0 & $\pm$8.4 & $\pm$7.4 & $\pm$2.9 & $\pm$0.4 & $\pm$119.0 \\
 a2 &  0.36$\times$0.10 & -0.45  & 44.1    & 4084.4 & 63.6  & 9.2   & 15.5  & 11.4  & 7.6  & 1.4 & 2.1 & 1362.0\\
  & $\pm$0.025$\times$0.012 &  & $\pm$3.0  & $\pm$3.7 & $\pm$9.2 &   & $\pm$3.1 & $\pm$2.7 & $\pm$2.7 &$\pm$0.5& $\pm$0.4 & $\pm$419.0 \\ 
 a3 &  0.40$\times$0.25 & 0.74   & 62.6    & 4053.7 & 59.1  & 10.6  & 29.6  & 28.7  & 8.9 & 3.7 & 4.1 &1032.0\\
  & $\pm$0.023$\times$0.014 &  & $\pm$2.9  & $\pm$4.0 & $\pm$6.0  &  & $\pm$2.1 & $\pm$6.2 & $\pm$2.3 &$\pm$1.2 & $\pm$ 0.3 & $\pm$236.0 \\
 a4 &  0.64$\times$0.19 & 0.13   & 54.1    & 4047.6 & 62.6  & 9.6   & 28.5  & 21.6  & 9.4  & 2.6 & 3.9 & 1321.0 \\
  & $\pm$0.054$\times$0.011 &  & $\pm$4.0  & $\pm$2.4 & $\pm$7.2  &  & $\pm$1.8 & $\pm$2.4 & $\pm$2.7 &$\pm$0.7& $\pm$0.2 & $\pm$171.0 \\
 a5 &  0.72$\times$0.44 & 3.38   & 123.9   & 4027.6 & 61.7  & 9.6   & 55.0  & 39.7  & 19.7  & 5.1 & 7.6 & 1850.0\\
  & $\pm$0.023$\times$0.022 &  & $\pm$4.3  & $\pm$1.2 & $\pm$5.6  &  & $\pm$1.5 & $\pm$6.3 & $\pm$4.5 & $\pm$1.5 & $\pm$0.2 & $\pm$221.0 \\
 b &   0.52$\times$0.38 & 5.09   & 95.7    & 4089.6 & 76.7  & 8.9   & 55.3  & 13.5  & 25.9  & 1.7 & 7.6 & 4084.0\\
  & $\pm$0.028$\times$0.026 &  & $\pm$4.4  & $\pm$1.2 & $\pm$4.4 &   & $\pm$2.0 & $\pm$1.9 &$\pm$3.6 &$\pm$0.5 &$\pm$0.3 & $\pm$598.0 \\
 c &   0.72$\times$0.21 & -32.46  & 76.1   & 4027.8 & 69.1  & 9.4   & 41.8 & 48.3   & 16.0 & 6.2 & 5.7 &866.0\\
  & $\pm$0.025$\times$0.007 &  & $\pm$2.1  & $\pm$1.9 & $\pm$5.0  &  & $\pm$2.8 & $\pm$5.9 & $\pm$2.8 &$\pm$1.7& $\pm$0.4 & $\pm$120.0 \\
 d\tablenotemark{a} & 0.94$\times$0.84 & 0.00 & 182.0 & 4067.3 & 87.3 & 30.2 & 69.4  & 13.0 & 66.6  & 1.7 & 9.6& 5320.0\\
  & $\pm$0.016$\times$0.014 &  & $\pm$2.4  & $\pm$1.7 & $\pm$4.8 & & $\pm$5.5  & $\pm$3.8 & $\pm$8.4 &$\pm$0.7 & $\pm$0.8 & $\pm$1603.0 \\
 
\hline
\end{tabular}\\
\tablenotetext{1}{Using the image in Fig. 3d, corresponding to a larger beam of 0.4\arcsec\ $\times 0.4$\arcsec.}
\tablecomments{The flux shown in this table is measured after the primary beam correction.\\
  Table columns are as follows: \\
  Col.\,1: clump number(as Fig.\,1a shown). \\
  Col.\,2: Major and minor axes by 2-d Gaussian fit.\\
  Col.\,3: Major axis position angle (PA; N to E).\\
  Col.\,4: The effective clump radius after a deconvolution with the ALMA beam.\\
  (R = 1.91$\sqrt{(\sigma_x\times\sigma_y)}$, where $\sigma$=FWHM/2.3548; \citealt{1987ApJ...319..730S})\\
  Col.\,5: The line center velocity from 1d Gaussian fit to the line profile within an elliptical aperture with radii of (major and minor axes (FWHM)).\\
  Col.\,6: The line velocity FWHM width from 1d Gaussian fit within an elliptical aperture with radii of (major and minor axes (FWHM)).\\
  Col.\,7: Clump peak surface brightness from the clump intensity map.\\
  Col.\,8: The CO\,(6-5) flux within an elliptical aperture with radii of (major and minor axes (FWHM)).\\
  Col.\,9: The continuum flux density within an elliptical aperture with radii of (major and minor axes (FWHM)).\\
  Col.\,10: Virial mass (see the text).\\
  Col.\,11: Molecular gas mass (estimated from the dust continuum, see the text).\\
  Col.\,12: Molecular gas mass (estimated from the CO flux, see the text).\\
  Col.\,13: Ratio of the total CO\,(6-5) flux to the 435$\,\mu$m dust continuum flux density.}
\end{table*}

\end{document}